 \pdfoutput=1
\documentclass{article}

\usepackage{moreverb}
\usepackage{booktabs}
\usepackage{amssymb}
\usepackage{amsmath}
\usepackage{rotating}
\usepackage{dsfont}
\usepackage{setspace}
\usepackage{multirow}
\usepackage{lscape}
\usepackage{tabularx}

\begin{document}
\title{Propensity score analysis with partially observed confounders: how should multiple
imputation be used?}
\author{Cl\'emence Leyrat $^{1}$, Shaun R. Seaman $^{2}$, Ian R. White $^{2}$,\\ Ian Douglas $^{3}$, Liam Smeeth $^{3}$, Joseph Kim $^{1,4}$,\\
Matthieu Resche-Rigon $^{5,6}$, James R. Carpenter $^{1,7}$ and Elizabeth J. Williamson$^{1,8}$}
\maketitle

\footnotesize
\noindent$^{1}$Department of Medical Statistics, London School of Hygiene and Tropical Medicine, London WC1E 7HT, UK\\
$^{2}$MRC Biostatistics Unit, Cambridge CB2 0SR, UK\\
$^{3}$Department of Epidemiology and Population Health, London School of Hygiene and Tropical Medicine, London WC1E 7HT, UK\\
$^{4}$IMS Health, Real-World Evidence Solutions, London N1 9JY , UK\\
$^{5}$SBIM Biostatistics and Medical information, Hôpital Saint-Louis, APHP, 75010 Paris, France\\
$^{6}$ECSTRA Team (Epidémiologie Clinique et Statistiques pour la Recherche en Santé), UMR 1153 INSERM, Université Paris Diderot, Sorbonne Paris Cité, 75010 Paris, France \\
$^{7}$MRC Clinical Trials Unit at UCL, London WC2B 6NH, UK\\
$^{8}$Farr Institute of Health Informatics, London NW1 2DA, UK\\~\\

Corresponding author: Cl\'emence Leyrat, London School of Hygiene and Tropical Medicine. 
Department of Medical Statistics.
Keppel Street, London WC1E 7HT. UK\\
clemence.leyrat@lshtm.ac.uk

\normalsize
\begin{abstract}
Inverse probability of treatment weighting (IPTW) is a popular propensity score (PS)-based approach to estimate
causal effects in observational studies at risk of confounding bias. A major issue when estimating the
PS is the presence of partially observed covariates. Multiple imputation (MI) is a natural 
approach to handle missing data on covariates, but its use in the PS context raises three important
questions: (i) should we apply Rubin's rules to the IPTW treatment effect estimates or
to the PS estimates themselves? (ii) does the outcome have to be included in the imputation model? (iii) how should we
estimate the variance of the IPTW estimator after MI?
We performed a simulation study focusing on the effect of a binary treatment on a binary 
outcome with three confounders (two of them partially observed). We used MI with chained equations to create complete datasets and compared
three ways of combining the results: combining treatment effect estimates (MIte); combining the 
PS across the imputed datasets (MIps); or combining the PS parameters and estimating the PS of the average covariates 
across the imputed datasets (MIpar). We also compared the performance of these methods to complete case (CC)
analysis and the missingness pattern (MP) approach, a method which uses a different PS model for each pattern of
missingness. We also studied empirically the consistency of these 3 MI estimators.
Under a missing at random (MAR) mechanism, CC and MP analyses were biased in most cases when estimating the 
marginal treatment effect, whereas MI approaches had good performance in reducing bias
as long as the outcome was included in the imputation model. However, only MIte was unbiased in all the studied scenarios and
Rubin's rules provided good variance estimates for MIte. In practice, the PS estimated in the MIte approach showed
good balancing properties, whereas the estimated PS in MIps and MIpar only balanced the observed part of the confounders.
Nevertheless, in some situations, MIpar removed a large part of the confounding bias in the treatment effect estimate. We propose
a variance estimator for MIpar, which has good statistical properties in our simulations. In conclusion, 
when using MI in the IPTW context, MIte with the outcome included in the imputation model is the preferred approach.
\end{abstract}~

\section{Introduction}
Data from observational studies provide useful information to address health-related questions
and notably estimate treatment effect in real settings \cite{concato_randomized_2000}. However, because individuals are not 
randomised, the study groups are often not comparable, which may lead to confounding bias
\cite{cochran_controlling_1973} if these studies are analyzed without appropriate adjustment for confounding. 
Propensity scores (PS) have been proposed 
as a means to recover balance between groups on observed confounders and so
obtain a consistent estimate of the causal treatment effect \cite{rosenbaum_central_1983}. 
The PS is defined as the individual's probability of receiving the treatment rather than the control given 
their baseline characteristics \cite{guo_propensity_2009}. One popular method to achieve covariate 
balance between treatment groups is to re-weight
individuals according to their PS value. This approach, known as inverse probability of treatment 
weighting (IPTW) \cite{hirano_estimation_2001} aims to emulate the sample that would have been observed in a randomised trial. In practice, a major issue 
when estimating the PS is the presence of partially observed confounders, as the PS cannot be estimated for individuals with at least one missing covariate value.
A simple solution is to perform the analysis using only data those individuals with complete records
 (\textit{i.e.} with complete baseline measurements), but such an approach
can lead to a biased estimate of the treatment effect if the missingness mechanism is associated with the outcome \cite{white_bias_2010}.
Another approach, proposed by Rosenbaum and Rubin \cite{rosenbaum_reducing_1984}, is to use missingness patterns (MP)
to estimate a generalised propensity score: individuals are classified according their pattern of missingness,
and then the PS is estimated within each pattern using the fully observed confounders for these individuals. This approach has
three drawbacks: it relies on strong assumptions, some confounders are ignored for some patterns and it needs to have a sample size large enough in each stratum 
to estimate the PS \cite{qu_propensity_2009}. A popular alternative to handle missing data is multiple imputation (MI).
MI is used to fill in missing confounders a given number of times, replacing the missing values by random draws from
the appropriate predictive distribution. The model parameters are estimated in the resulting imputed datasets and
then are combined using Rubin's rules to obtain the overall parameter estimates and their variance accounting for
the presence of missing data \cite{carpenter_multiple_2012}. Because the IPTW estimator is a two-step estimator 
(involving models for PS estimation and treatment effect estimation, respectively), Rubin's rules could be applied in several different ways.
First, the IPTW estimator could be applied to each of the imputed datasets and then the resulting estimates averaged (MIte). Second,
each individual's PS could be averaged over the imputed datasets and then these PSs used to calculate a single IPTW estimate (MIps). Third, 
the PS parameters and each individual's imputed confounders could be averaged over the imputed datasets and these used
to calculate a single IPTW estimate (MIpar). 
MI of partially observed confounders in the PS model has been used in applications
\cite{hayes_using_2008,mattei_estimating_2008}, and
studied through simulations for PS matching \cite{mitra_comparison_2012,hill_reducing_2004}. However, unresolved questions remain about how
to use MI in the context of IPTW. In particular, 
it is still unclear if combining the treatment effects across the imputed datasets outperforms the methods of combining the PSs,
as well as if the outcome must be included in the imputation model.

Moreover, little is known about the variance of the IPTW estimator when combining the PS parameters.
Thus, the aim of this work is to study the three different MI strategies described above to handle missing confounders for IPTW, and to compare them with
CC analysis and the MP approach.

This paper is organised as follows: we first present a motivating example looking at the effect of statins on short-term mortality after pneumonia in Section 2. A brief description of IPTW in complete data, its variance and its underlying assumptions is given in Section 3. Section 4 presents different strategies to handle partially missing confounders for PS analysis, focusing in particular on MI. We provide an approximately unbiased estimator of the variance for IPTW when combining the PS parameters. The consistency and balancing properties of the MI approaches are studied in Section 5. Section 6 and 7 present the methods and results of a simulation study assessing the performance of these methods
to estimate treatment effect with IPTW for binary outcomes. The application of these approaches to the statin motivating example is presented in Section 8, followed by a discussion in Section 9.

\section{Motivating example: effect of statin use on short-term mortality after pneumonia}
To illustrate the importance of an adequate handling of missing confounders for PS analysis,
we focused on a published study of the effect of statin use on short term mortality after pneumonia \cite{douglas_effect_2011}.
We utilised the THIN database, which consists of anonymised patient records from general practitioners (GPs) in the UK. 
As of the end of 2015, the database represented 3.5 million unique active patients, 
or approximately 6\% of the UK population. The database has been found to be broadly representative of the UK population, 
and the validity of recorded information has been established in previous studies \cite{lewis_validation_2007,blak_generalisability_2011}. 
Douglas \textit{et al} carried out an analysis of 9073 patients who had a pneumonia episode, of whom 1398 were under statin treatment
when pneumonia was diagnosed. In the statin group, 305 patients (21.8\%) died within 6 months, while 2839 (37.0\%) of the non-users 
died within 6 months. However, statin users and non users were very different in terms of characteristics,
in particular on characteristics associated with mortality. 

In Douglas \textit{et al}, propensity scores were used to recover balance between groups. However, three important potential
confounders were only partially observed: body mass index (BMI), smoking status and alcohol consumption,
with respectively 19.2\%, 6.2\% and 18.5\% of missing data. In the original analysis,  a missing indicator method was used. This approach is similar to the missingness pattern approach
described later in this paper.

\section{Inverse probability of treatment weighting}

\subsection{Propensity score estimation and assumptions}
\label{assumptions}
Propensity scores (PS) were first introduced by Rosenbaum and Rubin in the context of observational 
studies with confounding \cite{rosenbaum_central_1983}. The PS has become a major tool in causal inference to
estimate the causal effect of a binary treatment $Z$ ($Z=1$ if treated, $Z=0$ otherwise). Formally, PS is the individual's probability
of receiving the treatment conditional on the individual's characteristics. The PS is usually estimated from the data using a logistic regression
model which predicts each individual's probability of receiving the treatment from their baseline characteristics \cite{dagostino_propensity_1998}:
\begin{equation}
\hat{e}_i=\frac{exp(\boldsymbol{x}_i^T \hat{\boldsymbol{\alpha}})}{1+exp(\boldsymbol{x}_i^T \hat{\boldsymbol{\alpha}})},
\label{PS}
\end{equation}
where $\hat{e}_i$ is the estimated PS for individual $i$ $(i=1,...,n)$, $\boldsymbol{x}_i=(1,...,X_{pi})$
the vector of $p$ observed baseline confounders and the intercept for individual $i$ and $\hat{\boldsymbol{\alpha}}$ is the (p+1) vector of
maximum likelihood estimates for the PS parameters. \\

The PS approach can be viewed within the Neyman-Rubin counterfactual framework for causal inference \cite{rubin_estimating_1974}. In this framework,
the causal effect of the treatment is defined as the contrast of the two  potential outcomes (counterfactuals) $Y^{Z=0}$ and $Y^{Z=1}$,
which are the outcomes that would have been observed if an individual had been not treated and treated, respectively. Three assumptions 
are needed to consistently estimate the causal effect of a treatment, and thus needed for PS analysis:
\begin{enumerate}
\item Positivity assumption: each individual has a non-null probability of receiving either treatment \cite{cole_constructing_2008}
\item Stable unit treatment value assumption (SUTVA): each individual has only one possible potential 
outcome value for each treatment \cite{rubin_comment:_1986}. This assumption can be divided in two parts \cite{schwartz_extending_2012}:
\begin{itemize}
\item each of the two potential outcomes is always the same for a given individual, whatever the conditions in which the 
treatment has been received, \textit{i.e.} $Y=Y^z$ if $Z=z$, where $Y$ is the observed outcome. In other words, the treatment has the same effect on
the individual's outcome regardless of how the individual came to be treated. This assumption is also called consistency
\item the two potential outcome values for an individual are not affected by the treatment received by other individuals
\end{itemize}
\item Strongly ignorable treatment assignment (SITA): treatment allocation and the potential outcomes
are conditionally independent given the confounders, \textit{i.e.} $(Y^{Z=0},Y^{Z=1}) \perp Z|\boldsymbol{x}$. This
implies that there are no unmeasured confounders.
\end{enumerate}

The key property of the PS is that it is a balancing score. That is, if these assumptions are valid and the PS 
model is correctly specified (the functional form of the PS is correct), 
the variables included in the PS model are
balanced between treatment groups at any level of the PS. In other words, individuals with close
PS values have similar distribution of their characteristics. This balancing property
of the PS and the three assumptions lead to the consistency of PS-based estimators. Initially, Rosenbaum and Rubin
proposed three different PS-based approaches to estimate causal effects \cite{rosenbaum_central_1983}: PS matching, subclassification 
(also known as stratification) and covariate adjustment. Although PS matching is the most common 
approach used nowadays, two major limitations exist: matching often discards a substantial number of individual from the analysis \cite{austin_comparison_2013}
and variance estimation after PS matching is not straightforward \cite{an_bayesian_2010}. The two other approaches have
drawbacks as well: residual bias due to heterogeneity within strata can remain with subclassification \cite{rosenbaum_reducing_1984},
whereas covariate adjustment can be biased in some circumstances \cite{hade_bias_2013}. In addition, covariate adjustment provides a conditional,
rather than a marginal effect. Thus, we focus on a fourth PS-based approach \cite{rosenbaum_model-based_1987}:
inverse probability of treatment weighting (IPTW).

\subsection{IPTW estimator and its variance for complete data}
IPTW aims to create a pseudo-population similar to a randomised trial by re-weighting the individuals
according to the inverse of their probability of receiving the treatment they actually received (\textit{i.e.} $\hat{e}_i^{-1}$
for treated individuals and $(1-\hat{e}_i)^{-1}$ for untreated individuals). Thus, the IPTW estimators for the marginal
proportions for a binary outcome $Y$, $\mu_1$ and $\mu_0$, among the treated and the untreated are \cite{williamson_variance_2014}:
\begin{equation}
\hat{\mu}_1=\left( \sum_{i=1}^{n} \frac{Y_i Z_i}{\hat{e}_i}\right) \left( \sum_{i=1}^{n} \frac{Z_i}{\hat{e}_i} \right)^{-1},\mbox{~~~~~}
\hat{\mu}_0=\left( \sum_{i=1}^{n} \frac{Y_i (1- Z_i)}{1-\hat{e}_i}\right) \left( \sum_{i=1}^{n} \frac{1-Z_i}{1-\hat{e}_i} \right)^{-1}.
\label{IPTW}
\end{equation}
$Z_i$ is the treatment indicator for individual $i$ ($Z_i=1$ if treated, 0  otherwise) and $Y_i$ is their outcome value.
From these two marginal estimates, it is possible to estimate 
a relative risk $\left(\frac{\hat{\mu}_1}{\hat{\mu}_0}\right)$, an odds ratio $\left(\frac{\hat{\mu}_1/(1-\hat{\mu}_1)}{\hat{\mu}_0/(1-\hat{\mu}_0)}\right)$ 
or a risk difference $\left(\hat{\mu}_1-\hat{\mu}_0\right)$ for a binary outcome.\\

The IPTW estimator, as with other PS-based estimators, is a "two-step estimator": a first step is needed for the PS estimation and the second step is the 
treatment effect estimation. If the uncertainty linked to the PS estimation in the first step is not taken into account 
in a second step, the repeated sampling variance of the treatment effect will be overestimated and inference will be conservative
\cite{an_bayesian_2010}. Lunceford and Davidian \cite{lunceford_stratification_2004} and Williamson \textit{et al.} 
\cite{williamson_variance_2014} proposed a large-sample variance estimator for the IPTW treatment effect estimator in which a correction term 
including the variance/covariance matrix of the estimated PS parameters is applied.

\section{Handling missing data in propensity score analysis}

A major issue in PS estimation from model (\ref{PS}) is the presence of partially observed confounders,
which can lead to bias in the treatment effect estimate if ignored.
In this section, we describe five methods for applying IPTW on incomplete data. We assume the treatment status $Z$ and 
outcome $Y$ are fully observed.

\subsection{Complete case analysis}
Complete case analysis (or complete records analysis) is the most basic method in the
presence of missing data. The PS is estimated only within the subgroup of individuals
with observed values for all of the variables included in the PS model, and
 only these individuals contribute to the estimation of the treatment effect
\cite{hill_reducing_2004}. Although the complete case analysis
provides an unbiased estimate of the parameters of an outcome regression model when the missingness is independent of the outcome  \cite{white_bias_2010}, little
is known about complete case analysis for IPTW. Moreover,
 excluding individuals with missing confounders can reduce statistical power, 
particularly when the rate of missing data is high, because no use is made of partially
observed records.

\subsection{Missingness pattern approach}
The missingness pattern approach for PS analysis was first introduced by Rosenbaum
and Rubin \cite{rosenbaum_reducing_1984}. They defined a generalized PS $\hat{e}^*$ as
the probability of receiving the treatment given the observed confounders and the pattern of missing data.
In practice, the PS is estimated separately in each stratum defined by missingness patterns. For example,
if two variables $X$ and $W$ are partially observed, 4 different missingness patterns can exist:
 X and W are both observed, X is observed but W is missing, X is missing and W is observed,
or X and W are both missing. In each stratum, the PS is estimated from the confounders observed
in that stratum.
Under an extension of the SITA assumption, \textit{i.e.} when treatment allocation is independent of the potential outcomes
given the observed confounders and the missingness pattern, \textit{i.e.} $(Y^{Z=0},Y^{Z=1}) \perp Z|\boldsymbol{x}_{obs},r$
where $\mathbf{x_{obs}}$ is the observed component of $\mathbf{X}$ and $r$ is the missingness pattern, the generalized propensity score balances
the observed component of the partially observed confounders and the missingness indicators.
However, this generalized PS does not balance the unobserved component of the variables \cite{dagostino_estimating_2000}.

\subsection{Multiple imputation}

\subsubsection{Principles}~\\

Multiple imputation (MI) is a popular approach to handle missing data because of its flexibility and its efficiency in
a large variety of contexts and its implementation in many statistical software packages. The principle of
MI is to generate a set of plausible values for the missing variables by drawing from the posterior predictive distribution of
these variables given the observed data. $M$ complete datasets are created and analysed 
independently to produce estimates $\boldsymbol{\hat{\theta}_k}$, $(k=1,...,M)$ of $\boldsymbol{\theta}$ the vector of the parameters of interest
(\textit{eg.} regression coefficients) and estimates $\boldsymbol{W}_k$ of their associated variance matrix. 
Then $\boldsymbol{\hat{\theta}_k}$ and $\boldsymbol{W}_k$ , $(k=1,...,M)$  are combined across the $M$ imputed datasets.
Following Rubin's rules,  the overall estimate $\boldsymbol{\hat{\theta}_{MI}}$ of $\boldsymbol{\theta}$ and estimate of the variance
of $\boldsymbol{\hat{\theta}_{MI}}$, $\widehat{Var}(\boldsymbol{\hat{\theta}_{MI}})$,
 are estimated as follows \cite{carpenter_multiple_2012}:

\begin{equation}
\boldsymbol{\hat{\theta}}_{MI}=\frac{1}{M}\sum_{k=1}^M \boldsymbol{\hat{\theta}}_k, \mbox{~~~~~~~} \widehat{Var}(\boldsymbol{\hat{\theta}}_{MI})=\boldsymbol{W}+\left(1+\frac{1}{M}\right)\boldsymbol{B},
\label{Rubins}
\end {equation}
where $\boldsymbol{W}$ is the within-imputation variance-covariance matrix, which reflects the variability of the parameter estimates in each imputed dataset,
and $\boldsymbol{B}$ is the between-imputation variance matrix reflecting the variability in the estimates caused by the missing information.
These two components are defined as:
\begin{equation}
\boldsymbol{W}=\frac{1}{M}\sum_{k=1}^M \boldsymbol{W}_k, \mbox{~~~~~~~} \boldsymbol{B}=\frac{1}{M-1}\sum_{k=1}^M \left( \boldsymbol{\hat{\theta}}_k-\boldsymbol{\hat{\theta}}_{MI}\right)^2.
\end {equation}

Two approaches exist to impute datasets: joint modelling 
and chained equations. Joint modelling consists of imputing missing values
from a common parametric joint model for the complete data (often
a multivariate normal model) \cite{he_missing_2010}. However, when variables are of different types,
the existing classes of joint models may not be appropriate. The second approach,
chained equations (also known as fully conditional specification) is more flexible
in case of different variable types since a specific imputation model is
specified for each partially observed variable \cite{azur_multiple_2011}. 
Because of its flexibility, we focused on this approach in the context of PS. 

\subsubsection{How to apply Rubin's rules for PS analysis}~\\

Hayes and Groner used MI to estimate the PS from partially observed confounders
in a study looking at the effect of seat belt usage on injury severity \cite{hayes_using_2008}. However,
instead of combining information from their 15 imputed datasets, they randomly selected 
one complete record per individual and then estimated the PS. Because this approach does not
keep the entire benefit of MI, two approaches to combine information from the imputed datasets have been proposed
in the context of propensity score analyses: applying Rubin's rules on the treatment effect
or applying Rubin's rules on the PS itself. The former approach is the natural MI 
approach in which the parameter which is combined is also the parameter of interest: a PS
model is fitted on each imputed dataset, and the resulting PSs used to estimate a treatment effect.
Then formula (\ref{Rubins}) is applied to obtain an overall treatment effect and its variance
\cite{mitra_comparison_2012,hill_reducing_2004}. Seaman and White \cite{seaman_inverse_2014}
showed that Rubin's rule for estimating the variance performs well in practice in this setting, although theoretical justification for Rubin's rules
relies on the parameter of interest being estimated with maximum likelihood, which is not the case for the IPTW method. We will refer to this
approach as MIte hereafter. The second approach used in the literature consists of
combining each individual's PS across the imputed datasets (to obtain an average PS for each individual) and
then using these PSs to estimate a single treatment effect estimate \cite{mitra_comparison_2012,hill_reducing_2004}.
This method is called MIps hereafter. Because the PS is unlikely to follow a normal distribution, taking the average 
PS may not be appropriate. Thus, we 
propose a third combination method for PS analysis after MI, in which the PS parameter estimates $\hat{\boldsymbol{\alpha}}$
(ie.the regression coefficients of the PS model) are combined rather than the PS itself. Then
these parameters are used to estimate the PS corresponding to an individual's average imputed confounder values:
\begin{equation}
\hat{e}_i\left(\boldsymbol{\bar{x}}\right)=\frac{exp^{\boldsymbol{\bar{\alpha}\bar{x}_i}}}{1+exp^{\boldsymbol{\bar{\alpha}\bar{x}_i}}},
\end{equation}

\begin{figure}[!ht]
\includegraphics[scale=0.85]{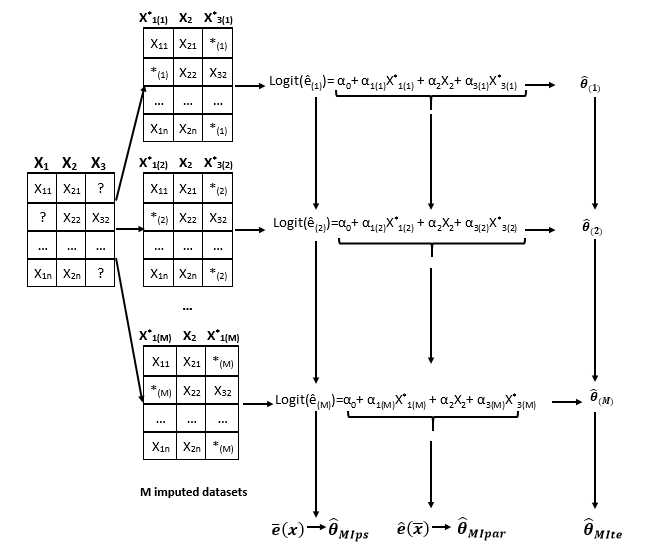}
\caption{The three approaches considered after multiple imputation (MI) of the partially observed
confounders}~\\  \footnotesize ? are missing values on the original dataset.
$*_{(k)}$, $(k=1,...,M)$ are imputed values in the $k^{th}$ imputed dataset. $\hat{\theta}_{(k)}$ and 
$\hat{e}_{(k)}$ are the estimated treatment effect and estimated propensity scores, respectively from the
$k^{th}$ imputed dataset , $(k=1,...,M)$. The MIte approach consists of pooling the M treatment effects 
estimated with IPTW  on each imputed dataset. MIps estimate is obtained by using the average PS across the
M imputed datasets in the IPTW estimator. Finally, the MIpar approach uses the PS of the average confounder
value across the M imputed dataset. The PS is estimated using the average PS parameters as regression coefficients.
\label{methods_MI}
\end{figure}

$(i=1,...,n)$ with $\boldsymbol{\bar{\alpha}}$ the $(p+1)$ vector of the average PS parameter values (for the p confounders and the intercept) and $\boldsymbol{\bar{x}_i}$
a $(p+1)$ vector of the average $p$ confounders  across imputed datasets for individual $i$. This method will be referred to as MIpar.
Whereas the MIps estimate of treatment effect is based on the average PS, $\bar{e}_i\left(\boldsymbol{x}_i\right)$,
the MIpar estimate is based on the PS of the average confounders, $e_i\left(\boldsymbol{\bar{x}}\right)$. 
The 3 MI approaches are illustrated in Figure \ref{methods_MI}.\\

Because the PS is obtained from $M$ imputations, the standard variance estimator for the IPTW treatment effect for MIps and MIpar is no longer valid
since it does not take into account the presence of missing data. A large-sample estimate
of the variance for MIpar, derived from \cite{williamson_variance_2014}, is detailed in Appendix 1.

The question of which PS strategy is preferable, as well as how to estimate the variance for MIps is still unresolved.
Moreover, the MIte and MIps methods have mainly been studied for PS matching \cite{mitra_comparison_2012,hill_reducing_2004}, suggesting a better reduction
of the bias observed in CC analysis with MIps than with MIte. Qu and Lipkovitch \cite{qu_propensity_2009} and Seaman and White \cite{seaman_inverse_2014}
assessed the performance of MIte additionally including the missingness pattern indicator  in the PS model, but they did not 
compare this approach to other combination rules after MI. For PS stratification 
Crowe \textit{et al.} \cite{crowe_comparison_2010} showed empirically that MIte has good statistical properties. However,
 little is known about performance of MIte and MIps in the context of IPTW. \\
To our knowledge,
the only simulation study comparing MIte and MIps with MI for IPTW did not take into account
the outcome in the imputation model \cite{mitra_comparison_2012}, which could explain the bias observed in their study
both for MIte and MIps. When using the PS approach, the fitting of the PS regression model is a first step to estimating 
the quantity which is of principal interest, namely the treatment effect. An attraction of this approach is that it enables 
the choice of how to adjust for confounding to be made without needing data on the outcome: choosing and fitting the PS model 
requires data only on covariates and treatment. This helps the user to avoid the temptation to search for a PS model that 
gives a significant treatment effect estimate. Intuition may therefore lead one to believe that imputation of missing 
covariates should also be done without using the outcome variable \cite{mitra_comparison_2012}. However, this intuition conflicts with advice to include 
the outcome when imputing missing covariates in a regression model whose parameters are the quantities of interest 
\cite{moons_using_2006}. In this setting, excluding the outcome causes the parameter estimates to be biased towards the null. 
The question remains, therefore, whether or not the outcome should be included when imputing missing covariates in the PS model.

\section {Balancing properties and consistency of IPTW estimator after MI}
Without missing data, Rosenbaum and Rubin showed that the 
PS is a balancing score \cite{rosenbaum_central_1983}. A balancing score $b(\boldsymbol{x})$ is defined as 
a function of the observed confounders $\boldsymbol{x}$ such that the conditional distribution of $\boldsymbol{x}$ given $b(\boldsymbol{x})$ is the same for $Z=0$ and $Z=1$. Moreover, Rosenbaum and Rubin showed that any balancing score $b(\boldsymbol{x})$ is 'finer' than the true PS, that is $e(\boldsymbol{x})=f \left\{b(\boldsymbol{x}) \right\}$, for some function $f(.)$. The consistency of PS estimators comes from this balancing property. \\
Lunceford and Davidian \cite{lunceford_stratification_2004} studied theoretical properties of the IPTW estimator when data are complete including a proof of consistency of this estimator.
In this section, we study the consistency of the IPTW estimators
obtained from MIte, MIps and MIpar and how this relates to balancing properties of the PS models used in these approaches. We suppose hereafter that (i) the SITA assumption required for IPTW (see \ref{assumptions})
holds, (ii) the missing data are missing at random (MAR) and (iii) the imputation model is correctly specified. 
For simplicity, we consider here the estimation of $\theta = \mathbb{E} [Y^{Z=1}].$ 

\subsection{Combining the treatment effects after MI (MIte)}

Let $\mathbf{X}$, the vector of confounders, be split into fully observed and missing components, $\mathbf{X} = (\mathbf{X}_{obs}, \mathbf{X}_{miss})$. $\mathbf{X}_{m}^{(k)}$ is the imputed value of $\mathbf{X}_{miss}$ in the $k^{th}$ imputed dataset ($k=1,..,M$). 
We show (see Appendix 2a) that in each imputed dataset:
\begin{equation} \label{eq: expPS_MIte}
e(\mathbf{X}_{obs}, \mathbf{X}_{m}^{(k)}) = E[Z | \mathbf{X}_{obs}, \mathbf{X}_{m}^{(k)}]. 
\end{equation}

If $\mathbf{X}_{m}^{(k)}$ is imputed from the true model (\textit{i.e.} correctly specified at the 
true parameter values), we can also show (Appendix 2b) that a
SITA-type assumption holds in each imputed dataset, \textit{i.e.}:
\begin{align} \label{eq: condexch_MIte}
&Y^{Z=1} \perp Z \, |  \, \mathbf{X}_{obs}, \mathbf{X}_{m}^{(k)} \\
&Y^{Z=0} \perp Z \, |  \, \mathbf{X}_{obs}, \mathbf{X}_{m}^{(k)} \nonumber
\end{align}
These two assumptions are the imputed-data version of what Imbens calls weak unconfoundedness \cite{imbens_role_2000}. Note that we do not have the analogy of the usual, stronger, assumption 
\begin{align*} 
(Y^{z=1}, Y^{z=0}) \perp Z \, |  \, \mathbf{X}_{obs}, \mathbf{X}_{m}^{(k)},
\end{align*}
which requires the treatment to be independent of the set of potential outcomes. This is because our imputation 
model is a model for $\mathbf{X}_{miss} | Z=z, Y^{Z=z}, \mathbf{X}_{obs}$. The stronger assumption would require our 
imputation model to capture $\mathbf{X}_{miss} | Z=z, Y^{Z=0}, Y^{Z=1}, \mathbf{X}_{obs}$. However, it is important to note, as Imbens does, that the weak
unconfoundedness suffices to obtain unbiased estimates of the causal treatment effect.

\textbf{Balancing properties}: We have shown in Appendix 2 that 
\begin{align*}
\mathbf{X}_{obs} \perp Z \, | \, e(\mathbf{X}_{obs}, \mathbf{X}_{m}^{(k)}) \\
\mathbf{X}_{m}^{(k)} \perp Z \, | \, e(\mathbf{X}_{obs}, \mathbf{X}_{m}^{(k)}).
\end{align*}
Thus the true PS in each completed dataset balance both the unobserved and the imputed values of the confounders across treatment groups. 
This balancing property is what leads to consistency of the MIte estimator. In practice, the PS has to be estimated.

\textbf{Consistency}:
Seaman and White \cite{seaman_inverse_2014} proved that for an infinite number of imputations, the estimator 
obtained by combining the treatment effects after MI (MIte) is consistent. To understand how this consistency
relates to the SITA-type assumption above, it is helpful to consider the following expectation:
\begin{align}
E\left[ \frac{Y Z}{e(\mathbf{X}_{obs}, \mathbf{X}_{m}^{(k)})} \right] & = 
E \left[E \left[ \left. \frac{Y Z}{e(\mathbf{X}_{obs}, \mathbf{X}_{m}^{(k)})}  \right| \mathbf{X}_{obs}, \mathbf{X}_{m}^{(k)} \right]  \right] because Z=0,1 \nonumber \\ & = 
E \left[ \frac{ \ E [ Y^{Z=1} | \mathbf{X}_{obs}, \mathbf{X}_m^{(k)}  ] \ E [ Z | \mathbf{X}_{obs}, \mathbf{X}_m^{(k)} ]}{e(\mathbf{X}_{obs}, \mathbf{X}_m^{(k)})} \right] 
\label{eq: MIte_condexch} \\ & = E [E[ Y^{Z=1}  |  \mathbf{X}_{obs}, \mathbf{X}_m^{(k)}]] 
 \label{eq: MIte_expPS} \\ & = E [Y^{Z=1}] \nonumber \\ & = \theta \nonumber,
\end{align}
Step \ref{eq: MIte_condexch}  requires the SITA-type assumption \ref{eq: condexch_MIte}. Step \ref{eq: MIte_expPS} relies on PS in the $k^{th}$ imputed dataset
being equal to the probability of being treated given the observed and imputed part of the confounders (equation \ref{eq: expPS_MIte}).

\subsection{Combining the PS or the PS parameters after MI (MIps and MIpar)}
For PS methods, consistency comes from the ability of PS to balance confounders between groups. MIps and MIpar create a single
 overall PS used to estimate the treatment effect. Thus, consistency for these methods would rely on the ability of these overall PS
to balance both the observed and the missing parts of the confounders.
However, when combining the PS or the PS parameters, the overall PS used for the analysis is not a function of the observed confounders. Thus, the pooled PS (as estimated either in MIps or MIpar) 
is not 'finer' than the true PS according to Rosenbaum and Rubin's definition \cite{rosenbaum_central_1983} (\textit{i.e.} the true PS is not a function of the pooled PS). 
Consequently, it cannot be a balancing score. Thus we do not have 
$\mathbf{X}_{obs}, \mathbf{X}_{miss}\perp Z|(\bar{e})$ (with $\bar{e}$ the relevant pooled PS). Therefore, 
neither $\hat{\theta}_{MIps}$ nor $\hat{\theta}_{MIpar}$ are consistent estimators. We illustrate the 
lack of consistency with a counter example in Appendix 3.  We also discuss the balancing properties of the MP approach in Appendix 4.
However, in practice, the consequences of this inconsistency on the treatment effect estimate is not known.
So we performed a simulation study to assess the performance of these estimators under different scenarios.

\section{Simulation study}
The aim of this simulation study is to assess the performance of the three MI approaches, complete case analysis and missingness pattern approach
for IPTW when the outcome is binary.
\subsection{Data generation}

We generated datasets of sample size $n=2000$, reflecting an observational study comparing a treatment $Z=1$ to a control
treatment $Z=0$ on a fully observed binary outcome $Y$ with 3 measured confounders $\mathbf{X}=(X_1,X_2,X_3)$.
$X_1$ and $X_2$ were continuous and 
$X_3$ was binary. $X_2$ was fully observed whereas $X_1$ and $X_3$ were partially observed. We generated the data
 as follows:

\begin{itemize}
\item \textbf{Covariates:} The 3 confounders $\mathbf{X}=(X_1,X_2,X_3)$ are generated from a multivariate normal distribution
$\mathbf{X} \sim N_3(\mathbf{0},\mathbf{\Sigma})$, with $\Sigma_{ii}=1$ and $\Sigma_{ij}=\rho$ for $i\neq j$. 
$X_3$ is then dichotomised according to a threshold of 0 to obtain a prevalence of 0.5.
We considered two values for $\rho:~\rho=0.3$ and $\rho=0.6$
corresponding to a moderate and strong correlation between confounders.
After dichotomization, $corr(X_1,X_3)=corr(X_2,X_3)=0.24$ for $\rho=0.3$ and 0.48 for $\rho=0.6$.\\

\item \textbf{Treatment assignment} depends on
$\mathbf{X}$ according to the following model:
\begin{equation}
logit(p(Z=1|\mathbf{x}))=-1.15+0.7 x_1+0.6 x_2+0.6 x_3.
\end{equation}
These coefficients give
$E(Z)=p_Z=0.3$ and an important imbalance on confounders between treatment groups, as shown in Figure 1 of the Appendices.\\

\item \textbf{Binary outcome:} The outcome depends on the 3 confounders and the treatment received
according the following model:
\begin{equation}
logit(p(Y=1|Z,\mathbf{x}))=-1.5+0.5 x_1+0.5 x_2+0.3 x_3+\theta_c Z.
\label{outcome2}
\end{equation}
In this model, $\exp{\theta_c}$ is the conditional odds ratio (OR). We used the method
described in \cite{austin_performance_2007} to find the value of $\theta_c$ to obtain the desired relative risk (RR). 
In order to have a RR of 1 $\theta_c=0$, and to have a RR of 2, $\theta_c=1.221$ and 1.289 for $\rho=0.3$ and $\rho=0.6$,
respectively.

With the non-null value of $\theta_c$, the marginal OR is 2.894 and the risk difference is 0.236
when $\rho=0.3$; they are 2.949 and 0.243 when $\rho=0.6$. 
With this model, the observed prevalence of the outcome is $p_0\approx 20\%$ in the control group. For the treated,
 $p_1\approx 33\%$ under the situation of no treatment effect $(\theta_c=0)$,  and $58\%$ under a non-null treatment effect situation $(\theta_c\neq=0)$
when $\rho=0.3$, and 35\% and 63\%, respectively for $\rho=0.6$.
 \\

\item \textbf{Missingness mechanism:}
In this simulation study, we consider a missing at random (MAR) mechanism. The missingness of 
each partially observed covariate ($X_1$ and $X_3$) depends on the fully observed covariate $X_2$, 
the treatment received $Z$ and the outcome $Y$ and was the same for $X_1$ and $X_3$: 
\begin{equation}
logit(p(R_1=1|Z,x_1,x_2,x_3,y))=\gamma_0+ z+ x_2+ \gamma_Y y,
\label {miss1}
\end{equation}
\begin{equation}
logit(p(R_3=1|Z,x_1,x_2,x_3,y))=\gamma_0+z+  x_2+\gamma_Y y,
\label {miss3}
\end{equation}
where $R_1$ and $R_3$ are the missingness indicators for $X_1$ and $X_3$ (equal to 1 if the value is missing), respectively. 
We set two values for $\gamma_Y$: $\gamma_Y=(0,-0.4)$ to have a situation in which the outcome value is not a predictor of missingness
(following Hill \cite{hill_reducing_2004} or Mitra and Reiter \cite{mitra_comparison_2012})
and a situation in which the missingness depends on the outcome (as in Bartlett \textit{et al.}  \cite{bartlett_multiple_2015} or 
Seaman and White \cite{seaman_inverse_2014}). When $\gamma_Y=0,~ \gamma_0=-1.5$ and when $\gamma_Y=-0.4,~ \gamma_0=-1.3$
to ensure the same missingness rate of 30\% for $X_1$ and $X_3$ across scenarios. \\

\end{itemize}

Because we considered two values for the correlation between confounders, two values for the treatment effect and two values
for the association between missingness and the outcome, a full factorial design leads to 8 scenarios both for binary and continuous outcomes.
We also looked at the impact of the omission of the outcome in the imputation model in the 8 main scenarios.
For one scenario (RR=2, $\rho=0.6$ and the outcome predictor of missingness), we studied the impact of the number of
imputed datasets ($M=5$ and $M=20$), the sample size ($n=500$) and the missingness rate ($10\%$ and $60\%$) on the results.
We also generated 30 \% missing data on the outcome and the treatment indicator on this scenario assuming MCAR.
 
For each scenario, 5000 datasets were simulated. Simulations were performed in R and the \textit{mi} package
 was used for multiple imputation \cite{R_mi}, based on full conditional specification (FCS) to generate $M=10$ imputed datasets.\\

\subsection{Estimated parameters}

For each studied scenario, we estimated the absolute bias of the treatment effect estimator and its
variance estimator, the empirical variance, the coverage rate and the standardized differences of the confounders
after IPTW. In the absence of weighting, standardized differences are defined as:
\begin{equation}
\mbox{SDiff}=\frac{100 \times \left|\bar{X}_{1}-\bar{X}_{0}\right|}{\sqrt{\frac{\hat{s}^2_{1}+\hat{s}^2_{0}}{2}}},
\label{SD1}
\end{equation}
for continuous variables, with $\bar{X}_{0},~ \bar{X}_{1},~ \hat{s}^2_{0}$ and $\hat{s}^2_{1}$ denoting the average value for covariate and its estimated variance
in the control and treatment group, respectively. For the binary confounders, the SD is:
\begin{equation}
\mbox{SD}=\frac{100 \times \left|\hat{P}_1-\hat{P}_0\right|}{\sqrt{\frac{\hat{P}_0(1-\hat{P}_0)+\hat{P}_1(1-\hat{P}_1)}{2}}},
\label{SD2}
\end{equation}
with $\hat{P}_0$ and $\hat{P}_1$ denoting the
estimated proportion of the covariate in the control and treatment groups.
~\\

For PS weighting, standardized differences are calculated by replacing the unweighted means and variances 
in (\ref{SD1}) and (\ref{SD2}) by their weighted equivalents (weighted by inverse PS).
For the MIte approach, standardized differences were calculated using the PS estimated from each imputed dataset,
both to assess the balance on the originally simulated complete dataset (before imposing missingness) and on the given imputed dataset.
For MIps and MIpar, standardized differences were calculated using the pooled PS 
to assess balance on (i) the original dataset, (ii) on the average value of the confounders across the imputed datasets. For (ii)
we also calculated the standardized differences separately on the observed part of the confounders and the average imputed part.

\section{Results}

Because results were similar for the three measures 
of interest, we present the results for relative risks (RR) only in the main text, while results 
for odds ratios and risk differences are in the appendices.

\subsection{Bias}
The absolute bias of the log(RR) of the treatment, for $\rho=0.6$, is presented in Figure \ref{bias}. Since results 
for $\rho=0.3$ are similar, they are presented in the Appendices. \\

\textbf{Full data, CC and MP analyses}: As expected,
the IPTW estimator on the full data (before generating missingness for $X_1$ and $X_3$) is approximately unbiased and the 
complete case (CC) estimator is strongly biased in all scenarios 
except those where the outcome is not associated with missingness and there is no treatment effect.
The MP approach is always biased in the situations considered, with a bias which can be even stronger than 
that which is observed for the CC approach. The reason for this is an incorrect PS model specification in each pattern of missingness: in the strata 
in which a confounder is not observed, the confounder is omitted in the model. \\

\textbf{Multiple imputation}: First, the results show that the imputation model must include the outcome, 
even if the outcome is not a predictor of missingness. All 3 MI estimators are strongly biased in all scenarios
when the outcome is not included in the imputation model. Second, when the outcome is included in the imputation model, 
the 3 MI approaches lead to a decrease in bias relative to the crude analysis. However,
only the MIte approach  leads to an unbiased estimate 
in the 8 main scenarios. Combining the PS parameters to estimate the PS of the average confounders (MIpar)
performed better than combining the PS themselves, but both these approaches are slightly biased. \\

\subsection{Standardized differences between groups}

The bias observed for the MP, MIps and MIpar methods can be explained by a remaining imbalance
on the confounders between groups. Standardized differences for each covariate are in Table \ref{tab_SD}.
A covariate is usually considered adequately balanced if its standardized difference is $<10\%$. IPTW on the
full data achieved a very good balance between groups on the 3 confounders (standardized difference $<$ 5\% for each of the 3 confounders).
For the CC approach, groups were balanced but the bias occurs since excluded individuals are different 
from included individuals on confounding factors. This can be seen as a selection bias, in which the sample analysed
is not representative of the target population.
The PS obtained from the MP approaches balanced the observed part of the confounders, but not the unobserved
part. This means that within each pattern of missingness,
treated and untreated individuals are balanced for the confounders included in the PS model, but unbalanced
on the missing covariate because this covariate is an unmeasured confounder in the PS model. 
Thus, when the missingness rate increases, imbalances (and consequently, bias of the treatment effect estimate)
increase (see section \ref{miss_rate}).
Whereas the PS estimated from each imputed dataset (MIte) balanced both the observed and imputed part of the confounders 
on the given imputed dataset, the average PS (MIps), the PS of the average confounders (MIpar) and the 
PS estimated with MP recovered the balance only the observed part of the confounders, as for the MP approach.
Moreover, in the MIte approach, the balance achieved on the fully observed covariate ($X_2$) is similar to
the balance observed on full data, whereas the imbalance on this variable with MIps and MIpar is slightly 
higher. Finally, the PS estimated in each imputed dataset (MIte) balances observed and imputed confounders in the given imputed
dataset, but this PS did not balance confounders in the original dataset.

\subsection{Coverage rate and standard errors}
Figure \ref{coverage} displays the coverage rate for each method when  the outcome is included in the imputation model. Each boxplot represents the coverage distribution for the 8 main scenarios. 
Because the CC and MP approaches are strongly biased, their coverage rates are not relevant. The coverage rate
for the MIte approach is close to the nominal value of 95\%, confirming that Rubin's rules perform well in
this context provided that the within-imputation variance takes into account the uncertainty in PS estimation.  \\

Table \ref{tab_var} shows the mean estimates from different variance estimators for each method when the outcome is included in the 
imputation model. It is important to note that using a variance
estimator for IPTW without taking into account the PS estimation leads to an overestimated variance. For the analysis on full data and for MIte, the
corrected variances are close to the empirical variance, whereas a naive estimator tends to overestimate the variance for these approaches.
A variance formula accounting for the PS estimation has not yet been provided for the MP approach.
For MIps and MIpar, the uncorrected variance estimator whose results are displayed in the table incorporates the variability linked to PS estimation
but not the imputation procedure. The proposed variance estimator (Appendix 1)
performs well in our simulations. In the scenarios presented in Table \ref{tab_var}, the corrected variance is smaller than the
uncorrected variance because the within imputation variance of the PS parameters (reflecting the correlation between 
the confounders and treatment; the higher this is the larger gain in precision for IPTW) is higher
than the between imputation variance component (noise due to missing data). However, when the missingness rate increases, 
the corrected variance can be higher than the uncorrected, because of a larger heterogeneity between imputed datasets.

\subsection{Sample size}
Table 17 of the Appendices present the results of one scenario with a non-null treatment effect
with a smaller sample size ($n=500$). Results were similar in terms of bias for $n=500$ and $n=2000$. Because the variance
estimator for IPTW has been developed for large samples, we observed slightly underestimated variances for 
the full data analysis, MIps and MIparam. This underestimation is more pronounced in the CC analysis because 
the sample for the analysis is even smaller (269 on average when $n=500$).

\subsection{Missingness rate}
\label{miss_rate}
Figure \ref{bias_miss} shows the bias when 10\% or 60\% of each partially observed covariate is missing. Full
results are presented in Appendix 6.3.
For a low missingness rate,  the CC and MP approaches are still biased but the 3 MI approaches corrected the bias. 
For a missingness rate of about 60\% for each covariate, only the MIte approach showed good performance in terms of bias reduction,
confirming the good statistical properties of this approach even with this large amount of missing data.

\subsection{Number of imputed datasets}
In our simulations, increasing the number of imputed datasets did not strongly impact the results
in terms of bias or variance (See Appendix 6.4).

\section{Application to the motivating example}
We applied CC analysis, the MP approach and the three MI strategies to estimate the effect of statin treatment on mortality 
after pneumonia from our motivating example dataset. For simplicity, we analysed the primary outcome, mortality within 6 months,
as a binary outcome, and estimated the corresponding relative risk and its 95\% confidence interval. For each approach, 
IPTW was used to account for the confounding. We focused the analysis on the 7158 patients without coronary heart disease.
The propensity score was estimated from a logistic regression 
modeling statin use as a function of the following covariates: age, sex, body mass index (BMI), alcohol consumption,
smoking status, diabetes,  cardiovascular disease  circulatory disease, heart failure,
dementia, cancer, hyperlipidaemia, hypertension and prescription of antipsychotics, hormone replacement therapy, antidepressants, 
steroids, nitrates,
beta-blockers,  diuretics, anticoagulants and use of antihypertensive drugs.
The imbalance between the study groups is illustrated in Figure \ref{THIN}.
Complete case (CC) analysis was conducted on the 5168 individuals with complete records. For the missingness pattern approach (MP), 
8 patterns were identified. However, some of these patterns were very rare. For instance, only 6 individuals had only the smoking status missing,
and only 8 had both smoking status and BMI missing. Thus, we considered only 4 groups:
\begin{itemize}
\item complete records (n=5168) for which all the covariates listed  above are included
\item individuals with only the alcohol consumption missing (n=455)
\item individuals with only BMI missing (n=575)
\item individuals with the smoking status missing (alone or in addition to BMI and alcohol consumption)
and individuals with both BMI and alcohol consumption missing (n=960)
\end{itemize}
For multiple imputation (MI), 10 imputed datasets were created. The imputation model included statin use, mortality and all the variables listed above.\\

The standardized differences estimated before weighting and after weighting by PS for CC, MP, MIte, MIps and MIpar are presented in Table \ref{tab_SDiff}.
The MP and the 3 MI approaches lead to a similar reduction in imbalance between groups on the observed variables as compared to the crude standardized differences.
Nevertheless, because of the poor overlap of the patients characteristics between groups (Figure \ref{THIN}), some covariates are still unbalanced even after MI.
However, for binary covariates, large standardized differences can occur even for slight imbalance when the prevalence is low.
Estimated RR are presented in Table \ref{res_THIN}. First, all approaches based on IPTW lead to a treatment effect estimate smaller than the unweighted treatment effect. The 3 MI approaches lead to similar RR and these were smaller than the RR obtained from CC and MP analyses. The small differences between the 3 MI approaches
in this example can be explained by a low rate of missing data and the fact that the 3 partially observed covariates were not strong confounders. 

\section{Discussion}

This paper aimed to address three main questions about multiple imputation in the context of IPTW: 
(i) does the outcome have to be included in the imputation model?
(ii) should we apply Rubin's rules on the IPTW treatment effect estimates or
on the PS estimates themselves?  (iii) how should we
estimate the variance of the IPTW estimator after MI?
First, results showed that the outcome must be included in the imputation model,
even if the outcome is not a predictor of missingness. This is well known in the context multivariable regression, 
but can be seen as counter intuitive in the PS paradigm since the PS model is built without reference to the outcome. 
The simulation results showed a bias in the 3 MI estimators when the outcome was omitted from the imputation model,
even when the outcome was not a predictor of missingness. This may explain the bias observed in Mitra and Reiter's study \cite{mitra_comparison_2012}
for PS matching.
Second, we showed that combining the treatment effects after MI (MIte approach) is the preferred MI strategy in terms of bias reduction
under a MAR mechanism.
This estimator is the only of the 3 MI estimator to be proven consistent and to provide good balancing properties.  Even though MIps and MIpar are not consistent estimators of the treatment effect,
they can reduce the bias observed for CC analysis, in particular when the rate of missing data is low. 
Combining the PS or the PS parameters has no clear advantage for IPTW, 
but may be useful in the context of PS matching: because it involves only one 
treatment effect estimation, it could provide computational advantage for large datasets. 
In addition, MIte for PS matching implies that the $M$ treatment effect estimates are estimated from
different matched sets, potentially of different sample sizes, leading to a more complex variance estimation because
of these different sample sizes. In our illustrative example, in which the missingness rate was moderate, the relative 
risks estimated with the 3 MI estimators were very similar.\\
Third, as
long as the uncertainty in the PS estimation is taken into account in the variance estimation \cite{williamson_variance_2014}, Rubin's rules perform
well for MIte, even for moderate sample size (n=500). For MIpar, the proposed variance approximation (Appendix 1) showed good performance in our simulation study.

The 3 MI approaches differ in terms of their balancing properties. We showed that whereas the PS estimated in each dataset in the MIte approach 
can balance confounders between groups in each imputed dataset, this is no longer true for MIps and MIpar. However, 
the best method to assess covariate balance after MI remains unknown. With MIte, the aim being to estimate a treatment effect from each 
dataset, we require balance between groups within each imputed dataset. In contrast, for MIps and MIpar, we need further investigation
to know if we should assess the balancing properties of the pooled PS on the average confounder values across the imputed dataset or on each dataset.\\

The MP approach, which is widely used in practice to handle missing data for PS analysis, revealed
poor performance under a MAR mechanism. This can be explained by the inability of the generalised propensity score
to balance the missing component of the confounders. Moreover, its application on our real life example was challenging because
of the sample size was not large enough within each missingness pattern to estimate the PS. 

This work has some limitations. We generated only 3 confounders in our simulation, whereas PS are
often built from a large number of confounders, despite the recent recommendation of parsimonious models \cite{brookhart_variable_2006}.  Moreover,
we studied only the common situation of log-linear relationships between the confounders
and the outcome and between the confounders and the treatment status. In the presence of interactions or quadratic terms 
in these two models, the specification of the imputation model can be less straightforward, requiring further efforts
to ensure the imputation model is compatible with the substantive (analysis) model \cite{bartlett_multiple_2015}. \\

In conclusion, for IPTW, multiple imputation followed by pooling of treatment effect estimates is the preferred approach amongst those studied
when data are missing at random, and the outcome must be included in the imputation model.\\~\\

\textbf{Funding}:
This work has been supported by the MRC (project grant MR/M013278/1).\\~\\

\textbf{Supplementary material}:
Complementary results are provided in Appendices

\bibliographystyle{Vancouver}
\bibliography{refs}

\clearpage

\begin{table}[htbp]
  \centering
  \caption{Standardized differences (in \%) after IPTW for each method for one scenario: RR=2, $\rho=0.6$, 
	outcome predictor of missingness and included in the imputation model. n=2000.}
    \begin{tabular}{rccc}
    \toprule
    \multicolumn{1}{c}{\textbf{Method}} & \textbf{X1} & \textbf{X2} & \textbf{X3} \\
    \toprule
    Crude (without IPTW) & 81.3  & 74.7  & 51.7 \\
    Full  & 4.6   & 4.6   & 2.4 \\
    CC (n=1074)   & 7.6   & 7.3   & 3.5 \\
    MP    &       &       &  \\
    \textit{Balance on full data} & 14.6  & 4.3   & 8.5 \\
    \textit{Balance on the observed part of the covariate} & 6.1   &   4.3    & 2.9 \\
    \textit{Balance on the missing part of the covariate} & 48.6  &  NA    & 28.3 \\
    MIte  &       &       &  \\
    \textit{Balance on full data} & 15.0  & 4.5   & 9.1 \\
    \textit{Balance on each imputed dataset } & 4.5   & 4.5     & 2.4 \\
    MIps  &       &       &  \\
    \textit{Balance on full data} & 15.9  & 5.5   & 10.7 \\
    \textit{Balance on the average imputed dataset} & 15.8  & 5.5     & 10.6 \\
    \textit{Balance on the observed part of the covariate} & 7.6   & 5.5     & 4.9 \\
    \textit{Balance on the imputed part of the covariate} & 58.1  & NA     & 36.9 \\
    MIpar &       &       &  \\
    \textit{Balance on full data} & 15.1  & 4.8   & 9.6 \\
    \textit{Balance on the average imputed dataset} & 14.7  & 4.8     & 9.7 \\
    \textit{Balance on the observed part of the covariate} & 7.7   & 4.8     & 5.4 \\
    \textit{Balance on the imputed part of the covariate} & 52.5  & NA     & 34.3 \\
    \bottomrule
    \end{tabular}%
		~\\
\vspace{0.2cm}
	\footnotesize
CC: complete case; MP: missingness pattern; MIte:
treatment effects combined after multiple imputation; MIps: propensity scores
combined after multiple imputation; MIpar: propensity score parameters 
combined after multiple imputation. RR: relative risk. 
NA: not applicable because $X_2$ is fully observed.
  \label{tab_SD}%
\end{table}%

~\\
\begin{landscape}
\begin{table}[htbp]
  \centering
  \caption{Bias of the log(RR), its estimated variance and coverage rate for the 3 MI approaches according the sample size $n$ for one scenario
	(RR=2, $\rho=0.6$, outcome predictor of missingness and included in the imputation model).}
	\scriptsize
    \begin{tabular}{rcccccccccccccc}
		
       \toprule
          & \multicolumn{2}{c}{\textbf{Full }} & \multicolumn{2}{c}{\textbf{CC}} & \multicolumn{2}{c}{\textbf{MP}} & \multicolumn{2}{c}{\textbf{MIte}} & \multicolumn{2}{c}{\textbf{MIps}} & \multicolumn{2}{c}{\textbf{MIpar}} \\
    \toprule
    \textbf{} & \textit{n=500} & \textit{n=2000} & \textit{n=500} & \textit{n=2000} & \textit{n=500} & \textit{n=2000} & \textit{n=500} & \textit{n=2000} & \textit{n=500} & \textit{n=2000} & \textit{n=500} & \textit{n=2000} \\
    Bias  & 0.007 & 0.002 & 0.110 & 0.141 & 0.153 & 0.130 & 0.010 & 0.005 & 0.038 & 0.028 & 0.024 & 0.017 \\
    Variance & 0.022 & 0.006 & 0.050 & 0.014 & 0.029 & 0.008 & 0.026 & 0.007 & 0.022 & 0.006 & 0.023 & 0.006 \\
    Empirical variance & 0.024 & 0.006 & 0.059 & 0.014 & 0.027 & 0.007 & 0.025 & 0.006 & 0.024 & 0.006 & 0.025 & 0.006 \\
    Coverage rate  & 0.940 & 0.947 & 0.887 & 0.769 & 0.855 & 0.691 & 0.955 & 0.957 & 0.939 & 0.932 & 0.943 & 0.942 \\
    \bottomrule

    \end{tabular}%

		~\\
		\vspace{0.2cm}
	\footnotesize MIte:
treatment effects combined after multiple imputation; MIps: propensity scores
combined after multiple imputation; MIpar: propensity score parameters 
combined after multiple imputation.  
  \label{tab_n}%
\end{table}%

\clearpage

\begin{table}[htbp]
  \centering
  \caption{Uncorrected and corrected model-based variance and empirical variance of the treatment effect estimator for the 8 main scenarios.}
	\scriptsize
    \begin{tabular}{c|ccc|ccc|cc|ccc|ccc|cccccc}

		    \toprule
    \multicolumn{1}{c}{\multirow{2}[4]{*}{Scenario}} & \multicolumn{3}{c}{Full} & \multicolumn{3}{c}{CC} & \multicolumn{2}{c}{MP} & \multicolumn{3}{c}{MIte} & \multicolumn{3}{c}{MIps} & \multicolumn{3}{c}{MIpar} \\
 
    \multicolumn{1}{c}{} & $\hat{V}_{u}$   & $\hat{V}_{PS}$ & $\hat{V}_{e}$   & $\hat{V}_{u}$   & $\hat{V}_{PS}$  & $\hat{V}_{e}$   & $\hat{V}_{u}$   & $\hat{V}_{e}$   & $\hat{V}_{m}$   & $\hat{V}_{PS+m}$  & $\hat{V}_{e}$   & $\hat{V}_{PS}$   & $\hat{V}_{PS+m}$  & $\hat{V}_{e}$  & $\hat{V}_{PS}$   & $\hat{V}_{PS+m}$  & $\hat{V}_{e}$ \\
		\toprule
      RR=1, $\rho=0.3$, Y    & 0.010 & 0.009 & 0.009 & 0.029 & 0.025 & 0.026 & 0.011 &  0.009 & 0.011 & 0.010 & 0.009 & 0.009 & 0.008 & 0.009 & 0.010 & 0.009 & 0.009 \\
       RR=1, $\rho=0.3$   & 0.010 & 0.009 & 0.009 & 0.033 & 0.030 & 0.031 & 0.011 &  0.009 & 0.011 & 0.010 & 0.009 & 0.009 & 0.008 & 0.009 & 0.010 & 0.009 & 0.009 \\
       RR=2, $\rho=0.3$, Y    & 0.007 & 0.006 & 0.006 & 0.015 & 0.013 & 0.014 & 0.007 &  0.006 & 0.007 & 0.006 & 0.006 & 0.006 & 0.006 & 0.006 & 0.006 & 0.006 & 0.006 \\
        RR=2, $\rho=0.3$   & 0.007 & 0.006 & 0.006 & 0.017 & 0.015 & 0.016 & 0.007 &  0.006 & 0.007 & 0.006 & 0.006 & 0.006 & 0.006 & 0.006 & 0.006 & 0.006 & 0.006 \\
        RR=1, $\rho=0.6$, Y   & 0.011 & 0.009 & 0.009 & 0.032 & 0.027 & 0.029 & 0.012 &  0.010 & 0.012 & 0.009 & 0.009 & 0.010 & 0.009 & 0.009 & 0.010 & 0.009 & 0.009 \\
         RR=1, $\rho=0.6$  & 0.011 & 0.009 & 0.009 & 0.037 & 0.033 & 0.035 & 0.011 &  0.010 & 0.012 & 0.009 & 0.009 & 0.010 & 0.009 & 0.009 & 0.010 & 0.009 & 0.009 \\
         RR=2, $\rho=0.6$, Y  & 0.008 & 0.006 & 0.006 & 0.017 & 0.014 & 0.014 & 0.008 &  0.007 & 0.008 & 0.007 & 0.006 & 0.007 & 0.006 & 0.006 & 0.007 & 0.006 & 0.006 \\
         RR=2, $\rho=0.6$  & 0.008 & 0.006 & 0.006 & 0.019 & 0.016 & 0.017 & 0.008 &  0.007 & 0.008 & 0.007 & 0.006 & 0.007 & 0.006 & 0.006 & 0.007 & 0.006 & 0.006 \\
    \bottomrule

    \end{tabular}%
  \label{tab_var}%
	~\\~\\
	 \footnotesize 
	
	\begin{flushleft}
Numbers are mean estimates of the variances in the 8 main scenarios.	Y means the outcome is included in the imputation model. RR: relative risk; MP: missingness pattern; MIte:
treatment effects combined after multiple imputation; MIps: propensity scores
combined after multiple imputation; MIpar: propensity score parameters 
combined after multiple imputation.   \\
$\hat{V}_{u}$ Variance estimate not accounting for the uncertainty in PS estimation or missing data\\
$\hat{V}_{PS}$ Variance estimate accounting for the uncertainty in PS estimation (Williamson \textit{et. al} formula)\\
$\hat{V}_{m}$ Variance estimate accounting for missing data but not for the uncertainty in PS estimation \\
$\hat{V}_{PS+m}$ Variance estimate using Williamson's formula and accounting for the presence of missing data\\
$\hat{V}_{e}$ Empirical variance
	\end{flushleft}
	
\end{table}%
\end{landscape}
\clearpage
\begin{landscape}
\begin{table}[htbp]
  \centering
  \caption{Description and comparison of statin users and non users. n=7158.}
\scriptsize
    \begin{tabular}{rrccccccccc}
    \toprule
    \multicolumn{1}{c}{\multirow{2}[2]{*}{\textbf{Variable}}} & \multicolumn{1}{c}{\multirow{2}[2]{*}{\textbf{Missing (\%)}}} & \textbf{Statin users} & \multirow{2}[2]{*}{\textbf{Missing (\%)}} & \textbf{Non statin users} & \multicolumn{6}{c}{\textbf{Standardized difference (\%)}} \\
    \multicolumn{1}{c}{} & \multicolumn{1}{c}{} & \textbf{n=599} &       & \textbf{n=6559} & \textbf{Crude} & \textbf{CC*} & \textbf{MP} & \textbf{MIte} & \textbf{MIps} & \textbf{MIpar} \\
			\toprule
    \multicolumn{1}{l}{\textbf{Characteristics}} & \multicolumn{1}{l}{\textbf{}} & \textbf{} & \textbf{} & \textbf{} & \textbf{} & \textbf{} &       &       &       &  \\
	
    Age [mean (sd)] &       & 66.9 (10.7) &       & 69.8 (10.9) & 27.0  & 3.8   & 2.0   & 1.4   & 1.4   & 1.4 \\
    Male  &       & 322 (53.8) &       & 3173 (48.4) & 10.8  & 2.0   & 6.2   & 2.2   & 2.1   & 2.2 \\
    BMI  [mean (sd)] & \multicolumn{1}{c}{43 (7.2)} & 27.6 (5.9) & 1444 (22.0) & 25.8 (5.9) & 31.9  & 7.8   & 9.0   & 9.0   & 11.4  & 11.4 \\
    Drinkers & \multicolumn{1}{c}{67 (11.2)} & 98 (18.4) & 1334 (20.3) & 814 (15.6) & 7.6   & 2.1   & 0.3   & 2.3   & 2.9   & 3.0 \\
    Smokers & \multicolumn{1}{c}{7 (1.2)} & 256 (43.2) & 505 (7.7) & 2728 (45.1) & 3.7   & 1.7   & 1.5   & 2.8   & 3.0   & 3.0 \\
          &       &       &       &       &       &       &       &       &       &  \\
    \textbf{Medical history} & \textbf{} &       &       &       &       &       &       &       &       &  \\
    Diabetes &       & 243 (40.6) &       & 715 (10.9) & 72.1  & 5.0   & 7.7   & 7.1   & 7.2   & 7.1 \\
    Cardiovascular disease &       & 141 (23.5) &       & 651 (9.9) & 37.1  & 11.4  & 11.4  & 13.6  & 13.6  & 13.6 \\
    Circulatory disease &       & 426 (71.1) &       & 3471 (52.9) & 38.2  & 13.6  & 9.8   & 16.6  & 16.7  & 16.6 \\
    Heart failure &       & 51 (8.5) &       & 426 (6.5) & 7.7   & 11.6  & 6.2   & 12.8  & 12.8  & 12.8 \\
    Cancer &       & 37 (6.2) &       & 607 (9.2) & 11.5  & 2.1   & 0.4   & 0.4   & 0.0   & 0.1 \\
    Dementia &       & 6 (1.0) &       & 190 (2.9) & 13.7  & 7.3   & 13.0  & 11.6  & 11.6  & 11.6 \\
    Hypertension &       & 336 (56.1) &       & 1165 (17.8) & 52.1  & 13.3  & 21.5  & 18.7  & 18.7  & 18.7 \\
    Hyperlipidemia &       & 205 (34.2) &       & 182 (2.8) & 88.5  & 1.1   & 4.1   & 1.9   & 2.0   & 2.0 \\
          &       &       &       &       &       &       &       &       &       &  \\
    \textbf{Treatments} & \textbf{} &       &       &       &       &       &       &       &       &  \\
    Antidepressant &       & 108 (18.0) &       & 995 (15.2) & 7.7   & 1.7   & 5.9   & 0.3   & 0.1   & 0.1 \\
    Antipsychotic &       & 11 (1.8) &       & 340 (5.2) & 18.3  & 0.5   & 11.3  & 5.0   & 5.0   & 5.0 \\
    Hormone replacement therapy &       & 37 (6.2) &       & 277 (4.2) & 8.8   & 0.9   & 0.9   & 1.0   & 1.0   & 1.0 \\
    Steroid &       & 93 (15.5) &       & 1090 (16.6) & 3.0   & 1.0   & 2.2   & 0.4   & 0.3   & 0.3 \\
    Antihypertensive &       & 272 (45.4) &       & 1165 (17.8) & 62.3  & 12.6  & 27.5  & 18.0  & 17.8  & 17.9 \\
    Diuretics &       & 319 (53.3) &       & 2416 (36.8) & 33.4  & 14.3  & 19.8  & 15.8  & 15.9  & 15.9 \\
    Betablocker &       & 193 (32.2) &       & 1061 (16.2) & 38.1  & 11.4  & 7.2   & 13.8  & 13.8  & 13.8 \\
    Nitrate &       & 74 (12.4) &       & 334 (5.1) & 25.9  & 17.3  & 14.8  & 17.5  & 17.6  & 17.6 \\
    \bottomrule
    \end{tabular}%

  \label{tab_SDiff}%
	~\\* For CC analysis, n=5168 (503 statin users and 4665 non users).
\end{table}%

\end{landscape}

~\\
\begin{table}[htbp]
  \centering
  \caption{Estimate of the relative risk of mortality and its 95\% confidence interval for statin vs non statin users (motivating example). n=7158.}
   
    \begin{tabular}{ccc}
    \toprule
    \textbf{Method} & \textbf{$\widehat{RR}$} & \textbf{95\% CI($\widehat{RR}$)} \\
    \midrule
    Crude & 0.587 & [0.497;0.684] \\
    CC    & 0.702 & [0.534;0.924] \\
    MP    & 0.708 & [0.555;0.904] \\
    MIte  & 0.654 & [0.513;0.835] \\
    MIps  & 0.653 & [0.512;0.834] \\
    MIpar & 0.654 & [0.513;0.834] \\
    \bottomrule
    \end{tabular}%

  \label{res_THIN}%
	~\\ \vspace{0.2cm}
	
	\begin{flushleft}
	RR: relative risk. CC: complete case; MP: missingness pattern; MIte:
treatment effects combined after multiple imputation; MIps: propensity scores
combined after multiple imputation; MIpar: propensity score parameters 
combined after multiple imputation. RR: relative risk. 
	\end{flushleft}

\end{table}%

~\\
~\\

\pagebreak

~\\
\begin{figure}[!ht]
\includegraphics[scale=0.50]{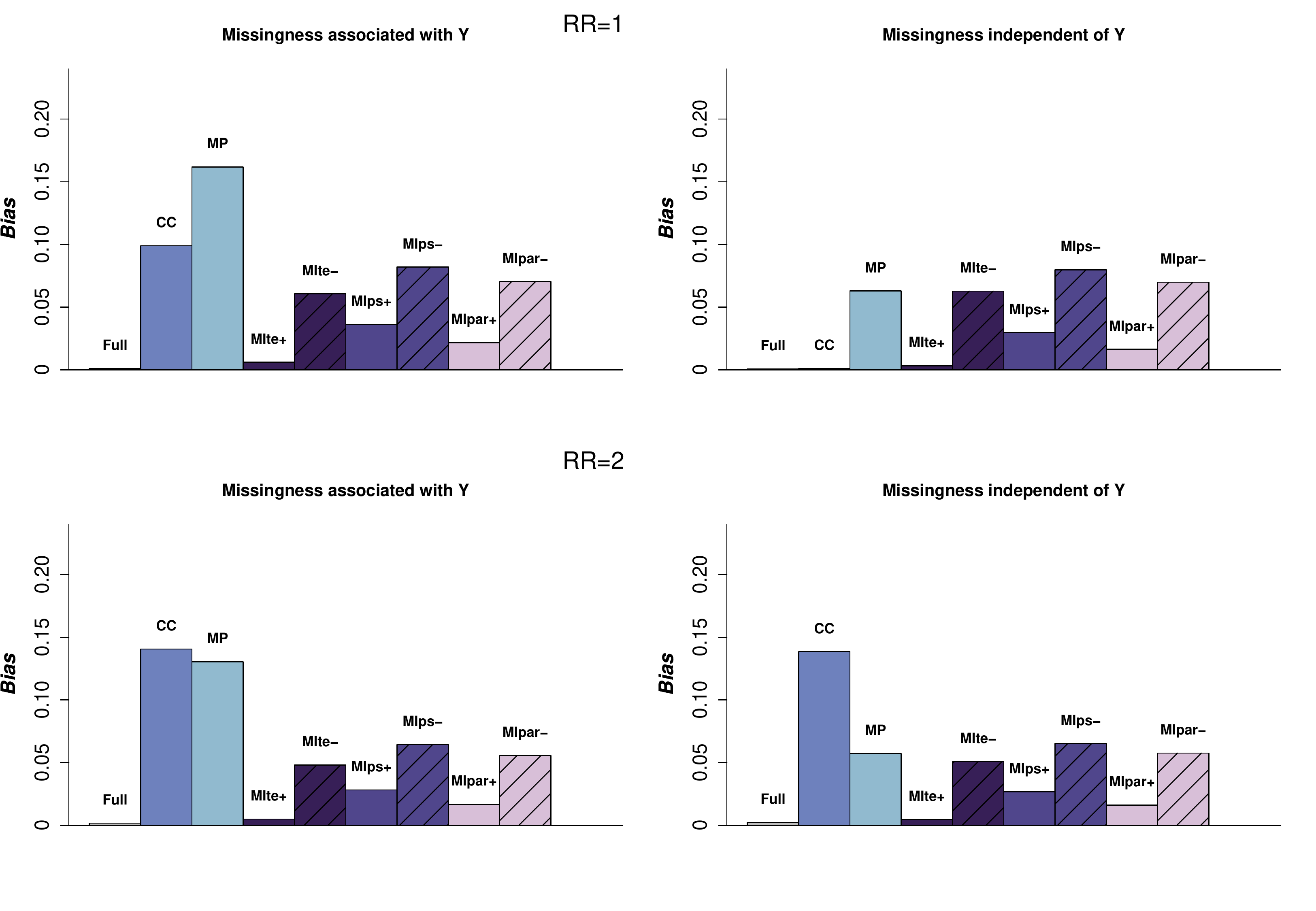}
\caption{Absolute value of the bias for the 8 main scenarios.}~\\  
\footnotesize CC: complete case; MP: missingness pattern; MIte:
treatment effects combined after multiple imputation; MIps: propensity scores
combined after multiple imputation; MIpar: propensity score parameters 
combined after multiple imputation. For the 3 MI approaches '+' means that the outcome is included in the imputation model,
'-' means that the outcome is not in the imputation model. RR: relative risk. 
\label{bias}
\end{figure}~\\

\pagebreak

~\\
\begin{figure}[!ht]
\centering
\includegraphics[scale=0.5]{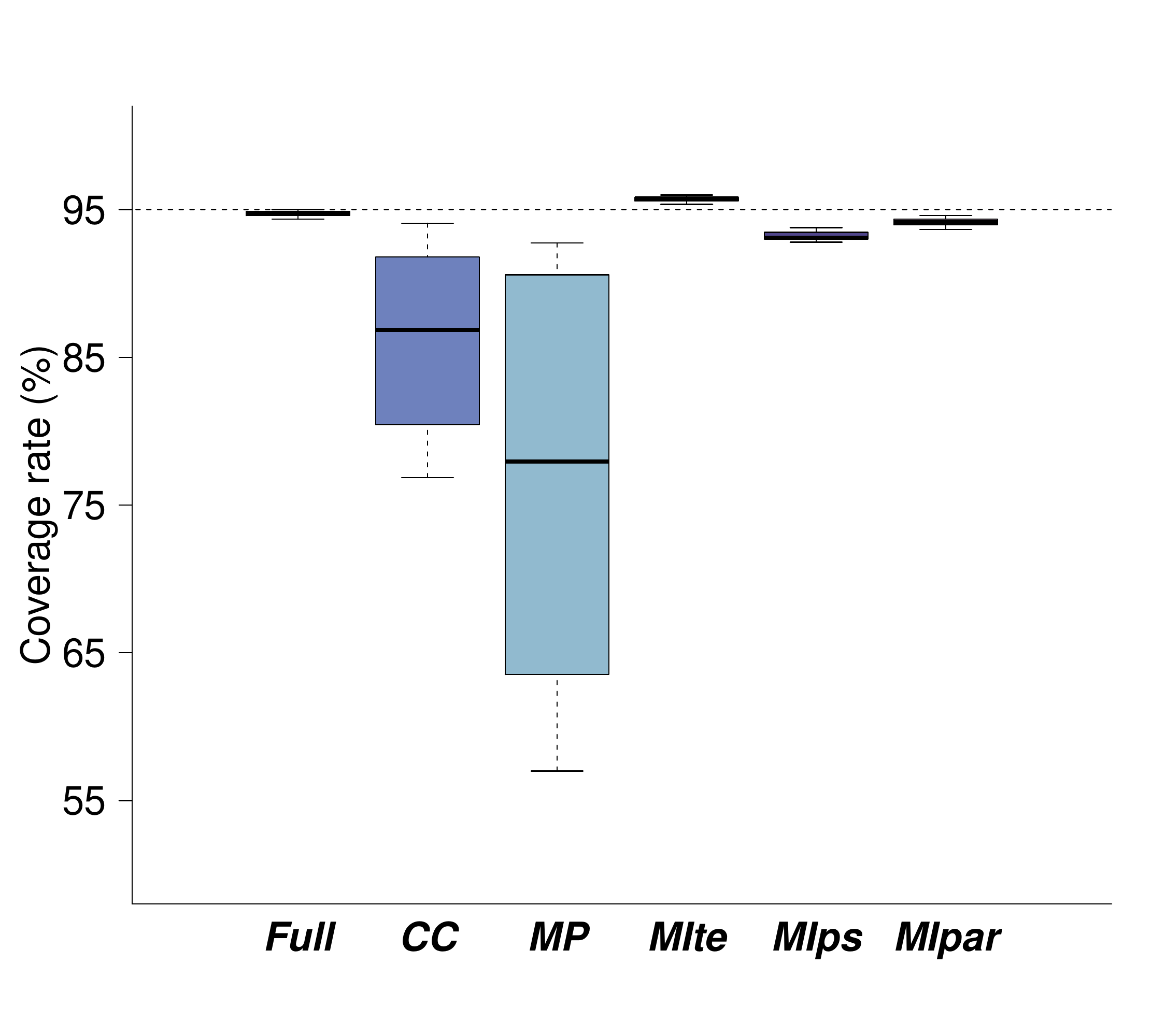}
\caption{Coverage rate of the 95\%CI for each method compared}~\\  
\footnotesize Results are pooled for the 8 main scenarios. 
CC: complete case; MP: missingness pattern; MIte:
treatment effects combined after multiple imputation; MIps: propensity scores
combined after multiple imputation; MIpar: propensity score parameters 
combined after multiple imputation. RR: relative risk.
\label{coverage}
\end{figure}~\\
\pagebreak

\begin{figure}[!ht]
\centering
\includegraphics[scale=0.55]{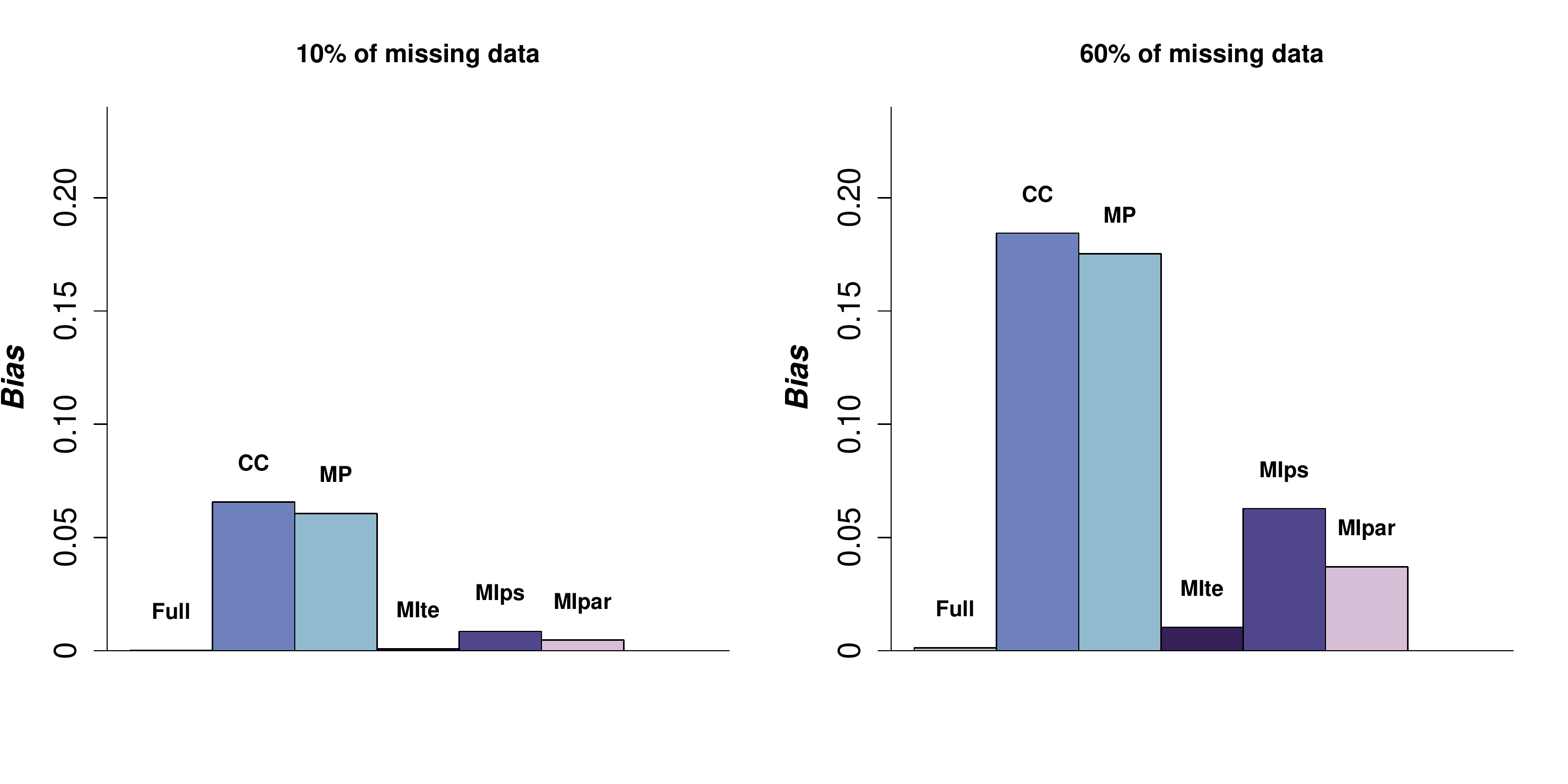}
\caption{Absolute value of the bias according to the missingness rate}~\\  
\footnotesize CC: complete case; MP: missingness pattern; MIte:
treatment effects combined after multiple imputation; MIps: propensity scores
combined after multiple imputation; MIpar: propensity score parameters 
combined after multiple imputation. RR: relative risk. 
\label{bias_miss}
\end{figure}~\\
\pagebreak

\begin{figure}[!ht]
\includegraphics[scale=0.75]{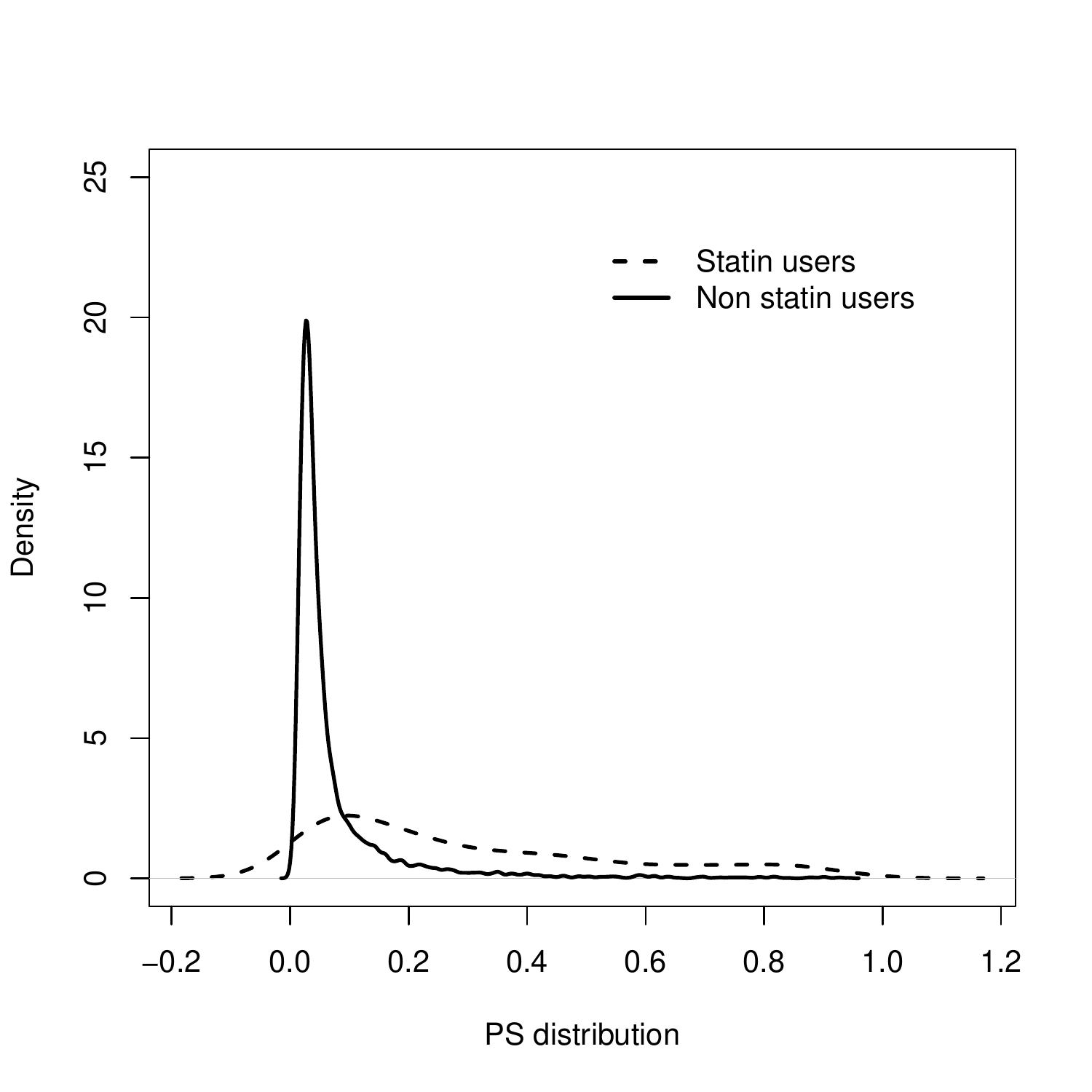}
\caption{Distribution of the propensity score estimated on complete cases for statin users and non users (n=5168).}~\\  
\label{THIN}
\end{figure}~\\

\clearpage

\Large{Appendix}
\normalsize

\section*{Appendix 1: Variance estimate for MIpar}
In this section, we provide a crude approximation to the variance for our MIpar estimator.
In MIpar, we take the average propensity score parameters across imputed datasets, $\overline{\boldsymbol\alpha}$, and the average of each covariate across imputed dataset for each individual, $\overline{\mathbf{X}}_i = (\mathbf{X}_{obs,i}, \sum_k \mathbf{X}_{m,i}^{(k)}/K)$ and use these to calculate the propensity score, $e(\overline{\mathbf{X}}_{i}; \overline{\boldsymbol\alpha})$. The MIpar treatment effect estimate is:
\begin{equation*}
\hat{\theta}_{MI_{par}}=  \frac{\sum_i \frac{Y_i Z_i}{e(\overline{\mathbf{X}}_{i}; \overline{\boldsymbol\alpha})} }{\sum_i \frac{Z_i}{e(\overline{\mathbf{X}}_{i}; \overline{\boldsymbol\alpha})}}  -   \frac{\sum_i \frac{Y_i (1-Z_i)}{(1-e(\overline{\mathbf{X}}_{i}; \overline{\boldsymbol\alpha}))} }{\sum_i \frac{(1-Z_i)}{(1-e(\overline{\mathbf{X}}_{i}; \overline{\boldsymbol\alpha}))}}  
\end{equation*}
If we let $I$ represent the imputation distribution, and $W$ the distribution of observed data $(Y, Z, \mathbf{X}_{obs})$, the variance of the treatment effect estimate is given by
\begin{equation} \label{eq: cond_var}
V[\hat{\theta}_{MI_{par}}] = \mathbb{E}_{W} \left[ V_{I|W} ( \hat{\theta}_{MI_{par}} ) \right]  + V_{W}\left( \mathbb{E}_{I|W} [ \hat{\theta}_{MI_{par}} ]\right)
\end{equation}
Suppose $\hat{\boldsymbol\alpha}^F$ denotes the propensity score parameters that would have been obtained in the full data, then 
\begin{equation*}
\mathbb{E}_{I|W} \, [ \ \hat{\theta}_{MI_{par}} \ ] = 
  \frac{\sum_i \frac{Y_i Z_i}{e(\overline{\mathbf{X}}_{i}; \hat{\boldsymbol\alpha}^F)} }{\sum_i \frac{Z_i}{e(\overline{\mathbf{X}}_{i}; \hat{\boldsymbol\alpha}^F)}}  -   \frac{\sum_i \frac{Y_i (1-Z_i)}{(1-e(\overline{\mathbf{X}}_{i}; \hat{\boldsymbol\alpha}^F))} }{\sum_i \frac{(1-Z_i)}{(1-e(\overline{\mathbf{X}}_{i}; \hat{\boldsymbol\alpha}^F))}}  + r_n
\end{equation*}
with $r_n \overset{p}{\to} 0$ as $n \to \infty$.  This is simply the full-data IPTW estimate, with the propensity scores evaluated at $\overline{\mathbf{X}}$ (and at the full data parameter estimates) rather than $\mathbf{X}$. Following Williamson \textit{et al.}, the same calculation shows that the large-sample variance of this full-data estimate, evaluated at $\overline{\mathbf{X}}$ rather than $\mathbf{X}$, is given by:
\begin{equation*}
V_{W} \left( \, \mathbb{E}_{I|W} \, [ \ \hat{\theta}_{MI} \ ]  \, \right) =  V_{un} - \mathbf{v}^T \mathbf{C}_{\boldsymbol\alpha} \mathbf{v} + \boldsymbol\epsilon
\end{equation*}

where
$V_{un}$ is the uncorrected variance estimate, that is considering that the PS is a true value rather than an estimate, which can be estimated by:
\[ \hat{V}_{un}=\frac{K_{1}^2}{n\hat{w_1}^2} \sum_{i=1}^n \frac{(Y_i-\hat{\mu}_1)^2 Z_i}{\hat{e}_i^2}+
\frac{K_{0}^2}{n\hat{w_0}^2} \sum_{i=1}^n \frac{(Y_i-\hat{\mu}_0)^2 (1-Z_i)}{(1-\hat{e}_i)^2},
\] and
\[ \mathbf{\hat{v}}=\frac{K_{1}}{n\hat{w_1}} \sum_{i=1}^n \frac{\mathbf{\bar{x}}_i(Y_i-\hat{\mu}_1) Z_i (1-\hat{e}_i)}{\hat{e}_i}+
\frac{K_{0}}{n\hat{w_0}} \sum_{i=1}^n \frac{\mathbf{\bar{x}}_i(Y_i-\hat{\mu}_1) (1-Z_i) \hat{e}_i}{(1-\hat{e}_i)}.
\]
$\hat{\mu}_1$ and $\hat{\mu}_0$ are the marginal (weighted) means in the treated and control group,   and 
$\hat{w}_1$ and $\hat{w}_0$ the average estimated weights for IPTW in the treated and control group, respectively,
and $\hat{e}_i=e(\mathbf{\bar{x}}_i; \mathbf{\hat{\alpha}})$. Finally,
the quantities $K_{0}$ and $K_{1}$ depend on the measure of interest for the treatment effect. For a difference in mean or a risk difference,
$K_0=K_1=1$. For a relative risk, $K_0=\hat{\mu_0}^{-1}$ and $K_1=\hat{\mu_1}^{-1}$, and for an odds ratio, 
$K_0=\left\{\hat{\mu_0}(1-\hat{\mu_0})\right\}^{-1}$ and $K_1=\left\{\hat{\mu_1}(1-\hat{\mu_1})\right\}^{-1}$.

$C_{\alpha} $ is $n^2\times$the variance/covariance matrix of the estimated PS parameters $\hat{\boldsymbol{\alpha}}$ in the full data, which we estimate by the within-imputation matrix $\mathbf{W}$.  The additional part of the variance is $\boldsymbol\epsilon = 2\mathbf{v} \mathbf{C}_{\boldsymbol\alpha} \boldsymbol\epsilon^*$, where  
\begin{align*}
\boldsymbol\epsilon^* &= \frac{1}{nE[Z/e(\mathbf{\bar{x}}; \mathbf{\alpha}^F)]} \mathbb{E} \left[ \mathbf{x}^\top (Y - \mu_1^F) Z  \left\{\frac{(e(\mathbf{x}; \boldsymbol\alpha^F) - e(\overline{\mathbf{x}}; \boldsymbol\alpha^F)}{e(\overline{\mathbf{x}}; \boldsymbol\alpha^F)}\right\} \right] \\
&+ \frac{1}{nE[(1-Z)/(1-e(\mathbf{\bar{x}}; \mathbf{\alpha}^F))]} \mathbb{E} \left[  \mathbf{x}^\top (Y - \mu_0^F) (1- Z)  \left\{\frac{(1 - e(\mathbf{x}; \boldsymbol\alpha^F)) - (1 -  e(\overline{\mathbf{x}}; \boldsymbol\alpha^F)))}{1 - e(\overline{\mathbf{x}}; \boldsymbol\alpha^F)} \right\} \right] 
\end{align*}
This is the only part of the variance which involves the full (unobserved) data. The magnitude of this term is driven partly by the difference between the propensity score evaluated at the average covariates across imputed datasets and the true (possibly unobserved) covariate values. For complete cases, therefore, this term is 0. 
We could attempt to estimate this component using the imputed datasets. However, we take a pragmatic approach and assume that pooling of the propensity scores is undertaken due to a desire to avoid the need to retain all $K$ imputed datasets. Therefore, we ignore this term in the simulation study and evaluate the performance of the estimator without this term. Thus, we assume that approximately
\begin{equation} \label{eq: cond_var 1}
\widehat{V}_{W} \left( \, \mathbb{E}_{I|W} \, [ \ \hat{\theta}_{MI} \ ]  \, \right) = \hat{V}_{un} - \hat{\mathbf{v}}^T \mathbf{W} \hat{\mathbf{v}} 
\end{equation}
This is our estimator for the second term in \ref{eq: cond_var}. For the first term, we begin by noting that,
 conditional on the observed data $(Y, Z, \mathbf{X}_{obs})$, we have approximately,
\begin{equation*}
V_{I|W} ( \hat{\theta}_{MI_{par}} )  = \left(\frac{\partial \hat{\theta}_{MI_{par}}}{\partial \overline{\boldsymbol\alpha}}\right)^T \, Cov_{I | W} \, (\overline{\boldsymbol\alpha}) \, \left(\frac{\partial \hat{\theta}_{MI_{par}}}{\partial \overline{\boldsymbol\alpha}}\right)
\end{equation*}
Noting that the average of the propensity score parameters across imputed datasets is also the standard MI estimate of these parameters,
\begin{equation*}
\widehat{Cov}(\overline{\boldsymbol\alpha})  = \left(1 + \frac{1}{M}\right) \ \mathbf{B}
\end{equation*}
where $\mathbf{B}$ is the between-imputation covariance matrix of the propensity score parameters. Differentiating the estimate above gives
\begin{align*}
 \frac{\partial \hat{\theta}_{MI}}{\partial \overline{\boldsymbol\alpha}} & = -
  \frac{\sum_i \left[ \frac{(Y - \hat{\mu}_1) Z \overline{\mathbf{x}} e(\overline{\mathbf{x}}; \overline{\boldsymbol\alpha}) (1-e(\overline{\mathbf{x}}; \overline{\boldsymbol\alpha}))}{(e(\overline{\mathbf{x}}; \overline{\boldsymbol\alpha}))^2} \right]  }{\sum_i\left[ \frac{Z}{e(\overline{\mathbf{x}}; \overline{\boldsymbol\alpha})}\right] } -
  \frac{\sum_i \left[ \frac{(Y - \hat{\mu}_0) (1-Z) \overline{\mathbf{x}} e(\overline{\mathbf{x}}; \overline{\boldsymbol\alpha}) (1-e(\overline{\mathbf{x}}; \overline{\boldsymbol\alpha}))}{(1-e(\overline{\mathbf{x}}; \overline{\boldsymbol\alpha}))^2} \right] }{\sum_i \left[ \frac{(1-Z)}{(1-e(\overline{\mathbf{x}}; \overline{\boldsymbol\alpha}))}\right]}   = - \hat{\mathbf{v}}
\end{align*}
Thus, we can estimate
\begin{equation} \label{eq: cond_var 2}
\widehat{\mathbb{E}} \left[ V_{I|W} ( \hat{\theta}_{MI} ) \right] = \left(1 + \frac{1}{M} \right)  \ \hat{\mathbf{v}}^T \mathbf{B} \hat{\mathbf{v}}
\end{equation}

Putting \eqref{eq: cond_var 1} and \eqref{eq: cond_var 2} into \eqref{eq: cond_var}, we have
\begin{align*}
\hat{V}( \hat{\theta}_{MI}) & = \hat{V}_{un} - \hat{\mathbf{v}}^T \left\{\mathbf{W}   - \left(1 + \frac{1}{M} \right)  \mathbf{B} 
 \right\} \hat{\mathbf{v}}
\end{align*}
Note that the two components of variability in the propensity score parameters - W and B - act in opposite directions on the variance of the treatment effect estimator. The between-imputation variance reflects noise due to the missing data, thus this adds to the overall variance of the treatment effect estimator. The within-imputation variance reflects the correlation between the confounders and treatment (ie covariate imbalance), which results in smaller variance of the treatment effect estimator (since the IPTW estimator gains precision by rebalancing imbalanced covariates).

\clearpage
\section*{Appendix 2: Assumptions required when pooling treatment effects}
\textbf{Notations} Let $\mathbf{X}$ the vector of confounders be split into observed and partially missing, $\mathbf{X} = (\mathbf{X}_{obs}, \mathbf{X}_{miss})$. $\mathbf{X}_{m}^{(k)}$ is the imputed value for $\mathbf{X}_{miss}$ in the $k^{th}$ imputed dataset ($k=1,..,M$) and $\mathbf{\alpha^{(k)}}$ the true propensity score parameters in the $k^{th}$ imputed dataset (with $\mathbf{\alpha^{(k)}}=\mathbf{\alpha}$, the overall true PS parameters, k=1,...,M). \\

\textbf{Appendix 2a:  Proof that $E[Z | \mathbf{X}_{obs}, \mathbf{X}_{m}^{(k)} ]  = e(\mathbf{X}_{obs}, \mathbf{X}_{m}^{(k)}; \boldsymbol\alpha^{(k)})$}\\
We have:
\begin{equation} \label{ap1}
E[Z | \mathbf{X}_{obs}, \mathbf{X}_{m}^{(k)}
 ] = \frac{Pr(\mathbf{X}_{m}^{(k)} | Z=1, \mathbf{X}_{obs}) Pr(Z=1 | \mathbf{X}_{obs} )}{\sum_{z=0,1} Pr(\mathbf{X}_m^{(k)} | Z=z, \mathbf{X}_{obs})  Pr(Z=z | \mathbf{X}_{obs} )}.
\end{equation}
We now express the probabilities $Pr(\mathbf{X}_m^{(k)} | Z=z, \mathbf{X}_{obs})$ in terms of the (unobserved) missing values:
\begin{align*}
Pr(\mathbf{X}_{m}^{(k)} | Z=z, \mathbf{X}_{obs}) & =  \sum_{y=0,1} Pr(\mathbf{X}_{m}^{(k)} | Z=z, \mathbf{X}_{obs}, Y=y) Pr(Y=y \, | \, Z=z, \mathbf{X}_{obs} )
\\ & = \sum_{y=0,1} Pr(\mathbf{X}_{miss} | Z=z, \mathbf{X}_{obs}, Y=y) Pr(Y=y \, | \, Z=z, \mathbf{X}_{obs} )
\\ & = Pr(\mathbf{X}_{miss} | Z=z, \mathbf{X}_{obs}),
\end{align*}if the values are imputed from the true distribution.
Substituting this back into \eqref{ap1} gives
\begin{align*}
E[Z | \mathbf{X}_{obs}, \mathbf{X}_{m}^{(k)}= \mathbf{x} ] & = E[Z | \mathbf{X}_{obs}, \mathbf{X}_{miss}=\mathbf{x}]  =  e(\mathbf{X}_{obs}, \mathbf{x}; \boldsymbol\alpha) = e(\mathbf{X}_{obs}, \mathbf{X}_{m}^{(k)} = \mathbf{x}; \boldsymbol\alpha^{(k)}).
\end{align*}

\clearpage
\textbf{Appendix 2b:  Proof that $Y^{z=1} \perp Z \, | \, \mathbf{X}_{obs}, \mathbf{X}_{m}^{(k)}$}\\
We have:
\begin{align}
&Pr(Z = 1 | Y^{z=1}=y_1, \mathbf{X}_{obs}, \mathbf{X}_{m}^{(k)} )  \nonumber \\ &= \frac{Pr(\mathbf{X}_{m}^{(k)}  | Z=1, Y^{z=1}=y_1, \mathbf{X}_{obs} ) Pr(Z=1  | Y^{z=1}=y_1,  \mathbf{X}_{obs} ) }{Pr(\mathbf{X}_{m}^{(k)}  | Y^{z=1}=y_1, \mathbf{X}_{obs} ) } \nonumber \\ & = \frac{Pr(\mathbf{X}_{miss}  | Z=1, Y^{z=1}=y_1, \mathbf{X}_{obs} ) Pr(Z=1  | Y^{z=1}=y_1,  \mathbf{X}_{obs} ) }{Pr(\mathbf{X}_{miss}  | Y_1=y_1, \mathbf{X}_{obs} )} \label{eq: MIte_y} \\ & = Pr(Z=1 | Y^{z=1}=y_1, \mathbf{X}_{obs}, \mathbf{X}_{miss}) \nonumber \\ & = Pr(Z=1 | \mathbf{X}_{obs}, \mathbf{X}_{miss}) \label{eq: MIte_sita}  \\ & = Pr(Z=1 | \mathbf{X}_{obs}, \mathbf{X}_{m}^{(k)}), \label{eq: MIte_prz}
\end{align}
where \eqref{eq: MIte_y} was reached by noting that $Y_1=Y$ when $Z=1$,   \eqref{eq: MIte_sita} is true by the SITA assumption for the full data (Assumption 1), and \eqref{eq: MIte_prz} was shown in Appendix 1a.

Proof that $Y^{z=0} \perp Z \, |  \, \mathbf{X}_{obs}, \mathbf{X}_{m}^{(k)}$ follows in a similar way.\\

\clearpage

\section*{Appendix 3: inconsistency of MIps and MIpar estimators: a counter example }
Neither $\hat{\theta}_{MIps}$ nor $\hat{\theta}_{MIpar}$ are consistent estimators. This can be shown by a simple counter-example. Suppose we have a single confounder, $X$, which is MAR depending on treatment $Z$ only:
\begin{align*}
&X \sim Bernouilli(0.5) \\
&Z \sim \begin{cases} 
Bernouilli(0.1) &\mbox{if} \ X=0  \\
Bernouilli(0.9) &\mbox{if} \ X=1  \\
\end{cases}  \\
&Y \sim \begin{cases} 
 Bernouilli(0.1) &\mbox{if} \ X=0 \ \mbox{and/or} \ Z=0  \\
 Bernouilli(0.9) &\mbox{if} \ X=Z=1  \\
\end{cases}  \\
&R \sim \begin{cases} 
1 &\mbox{if} \ Z=0   \\
Bernouilli(0.1) &\mbox{if} \ Z=1  \\
\end{cases} 
\end{align*}
$X_{obs} = {X:R=1}$ and $X_{miss} = {X:R=0}$. The treatment model corresponds to: $ln(e/(1-e)) = -2.1972246 + 4.3944492 X$, thus $\boldsymbol\alpha = (-2.1972246, 4.3944492)^T$. The true expected potential outcome is $\mathbb{E}[Y^{z=1}] = 0.5$. 

\textbf{Consistency when pooling propensity scores} When $Z=1$ and $Y=1$, the expected propensity score is
$\mathbb{E} [ e(X_{obs}, X^{(k)}_m; \boldsymbol\alpha)] = 
 \mathbb{E} \left[ \mathbb{E} [e(X, \boldsymbol\alpha) | X] \right]  = 0.89$. When $Z=1, Y=1$, $X$ can take values of $0, 1$ or missing, producing average (expected) propensity scores of 0.1, 0.9, and 0.89, respectively. Thus 
\begin{align*}
& \mathbb{E} \left[ \frac{YZ}{\mathbb{E} [ e(X_{obs}, X^{(k)}_m; \boldsymbol\alpha^t) ]} \right]  =   Pr(X=0, Z=1, Y=1, R=1)/0.1 \\ &+ Pr(X=1, Z=1, Y=1, R=1)/0.9 + Pr(Z=1, Y=1, R=0)/0.89 \\ &  = 0.465
\end{align*} 
So  $\mathbb{E} [ \frac{YZ}{\mathbb{E} \left[ e(X_{obs}, X^{(k)}_m; \boldsymbol\alpha)\right]} ] \neq \mathbb{E} [Y^{z=1}]$. Thus pooling the propensity scores will produce an inconsistent estimator here. 

\textbf{Consistency when pooling propensity score parameters} When $Y=1, Z=1$, the expected value of $X$ is $\mu_{bar{X}}\mathbb{E}[X_{miss} | Z=1, Y=1]  = Pr(X=1,Z=1,Y=1)/Pr(Z=1,Y=1) = 81/82$. The true propensity score evaluated at $\boldsymbol\alpha$ with $X=0, 1$ and $81/82$ is 0, 1, and $0.895$, respectively. Then,
\begin{align*}
&\mathbb{E} \left[ \frac{YZ}{e(X_{obs},\mu_{\overline{X}}; \alpha)} \right]  =   Pr(X=0, Z=1, Y=1, R=1)/0.1 \\ &+ Pr(X=1, Z=1, Y=1, R=1)/0.9 
+ Pr(Z=1, Y=1, R=0)/0.895 \\ & = 0.462
\end{align*} 
So  $\mathbb{E} \left[ \frac{YZ}{e(X_{obs},\mu_{\overline{X}}; \alpha)}\right]  \neq \mathbb{E} [Y^{z=1}]$. Thus pooling the propensity scores will produce an inconsistent estimator here also. 

\section*{Appendix 4: Consistency for the MP approach}

Let $e*$ be the generalized PS estimated from the MP approach.
Rosenbaum and Rubin showed that $\mathbf{X}_{obs} \perp Z|e*$. This implies 
that for each value of $e*$, the observed part of the confounders
and the frequency of missingness are balanced between groups. However,
they state that we do not have: 
$\mathbf{X} \perp Z|e*$.
Because we do not have this conditional independence, $e*$ is not a balancing score 
for $(\mathbf{X}_{obs}, \mathbf{X}_{miss})$ and thus the MP estimator of treatment effect is inconsistent,
without making further conditional independence assumptions.
\clearpage

\section*{Appendix 5: PS distribution in the simulation study}
\begin{figure}[!h]
\centering

\includegraphics[scale=0.6]{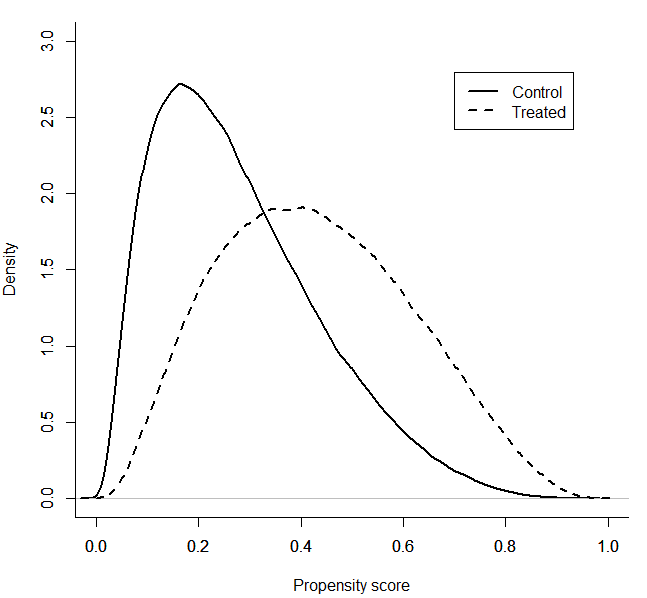}
\caption{Distribution of the propensity score in both groups in our simulation study.}~\\  \footnotesize 
These distributions are obtained from one dataset of size n=1000000.
\label{PS}
\end{figure}~\\

\clearpage

\section*{Appendix 6: Additional simulation results}
\subsection*{6.1: Full results for a binary outcome}
\begin{table}[htbp]
  \centering
  \caption{Scenario 1: RR=1, $\rho=0.3$ and outcome predictor of missingness and included in the imputation model.}
	\footnotesize
    \begin{tabular}{rcccccccccccccccc}
    \toprule
          & \textbf{Crude} & \textbf{Full } & \textbf{CC*} & \textbf{MP} & \textbf{MIte} & \textbf{MIps} & \textbf{MIpar} \\
    \toprule
    \textbf{RR} &       &       &       &       &       &       &  \\
    Bias  & 0.538 & 0.000 & 0.092 & 0.182 & 0.003 & 0.036 & 0.021 \\
    Variance  & 0.006 & 0.009 & 0.025 & 0.011 & 0.010 & 0.008 & 0.009 \\
    Empirical variance  & 0.006 & 0.009 & 0.026 & 0.009 & 0.009 & 0.009 & 0.009 \\
    Coverage rate & 0.000 & 0.947 & 0.898 & 0.570 & 0.959 & 0.929 & 0.944 \\
    \textbf{OR} &       &       &       &       &       &       &  \\
    Bias  & 0.722 & 0.000 & 0.121 & 0.239 & 0.004 & 0.048 & 0.028 \\
    Variance & 0.012 & 0.015 & 0.041 & 0.019 & 0.016 & 0.015 & 0.015 \\
    Empirical variance  & 0.012 & 0.015 & 0.042 & 0.016 & 0.015 & 0.015 & 0.015 \\
    Coverage rate  & 0.000 & 0.948 & 0.908 & 0.580 & 0.958 & 0.931 & 0.946 \\
    \textbf{Risk difference} &       &       &       &       &       &       &  \\
    Bias  & 0.136 & 0.000 & 0.022 & 0.043 & 0.001 & 0.009 & 0.005 \\
    Variance & 0.000 & 0.000 & 0.001 & 0.001 & 0.001 & 0.000 & 0.000 \\
    Empirical variance & 0.000 & 0.000 & 0.001 & 0.001 & 0.000 & 0.000 & 0.000 \\
    Coverage rate  & 0.000 & 0.949 & 0.928 & 0.602 & 0.960 & 0.935 & 0.947 \\
    \bottomrule

    \end{tabular}%
	\begin{flushleft}
	\footnotesize 
	* For complete case analysis, the average sample size is 1061.\\
	RR: relative risk; OR: odds ratio; CC: complete case; MP: missingness pattern;
	MIte:treatment effects pooled after multiple imputation; MIps: propensity scores
pooled after multiple imputation; MIpar: propensity score parameters 
pooled after multiple imputation.
	\end{flushleft}
\end{table}%

\begin{table}[htbp]
  \centering
  \caption{Scenario 2: RR=1, $\rho=0.3$ and outcome independent of missingness but included in the imputation model.}
	\footnotesize
    \begin{tabular}{rcccccccccccccccc}
    \toprule
          & \textbf{Crude} & \textbf{Full } & \textbf{CC*} & \textbf{MP} & \textbf{MIte} & \textbf{MIps} & \textbf{MIpar} \\
    \toprule
    \textbf{RR} &       &       &       &       &       &       &  \\
    Bias  & 0.539 & -0.001 & -0.007 & 0.082 & 0.002 & 0.030 & 0.016 \\
    Variance & 0.006 & 0.009 & 0.030 & 0.011 & 0.010 & 0.008 & 0.009 \\
    Empirical variance & 0.006 & 0.009 & 0.031 & 0.009 & 0.009 & 0.009 & 0.009 \\
    Coverage rate & 0.000 & 0.949 & 0.941 & 0.890 & 0.958 & 0.938 & 0.943 \\
    \textbf{OR} &       &       &       &       &       &       &  \\
    Bias  & 0.722 & 0.000 & -0.006 & 0.107 & 0.003 & 0.040 & 0.022 \\
    Variance & 0.012 & 0.015 & 0.044 & 0.018 & 0.016 & 0.015 & 0.015 \\
    Empirical variance & 0.012 & 0.015 & 0.046 & 0.016 & 0.015 & 0.015 & 0.015 \\
    Coverage rate & 0.000 & 0.949 & 0.943 & 0.895 & 0.958 & 0.939 & 0.945 \\
    \textbf{Risk difference} &       &       &       &       &       &       &  \\
    Bias  & 0.136 & 0.000 & 0.001 & 0.019 & 0.001 & 0.007 & 0.004 \\
    Variance & 0.000 & 0.000 & 0.001 & 0.001 & 0.001 & 0.000 & 0.000 \\
    Empirical variance & 0.000 & 0.000 & 0.001 & 0.001 & 0.000 & 0.000 & 0.000 \\
    Coverage rate  & 0.000 & 0.948 & 0.942 & 0.903 & 0.959 & 0.942 & 0.946 \\
    \bottomrule

    \end{tabular}%
	\begin{flushleft}
	\footnotesize
	* For complete case analysis, the average sample size is 1103.\\
	RR: relative risk; OR: odds ratio; CC: complete case; MP: missingness pattern;
	MIte:treatment effects pooled after multiple imputation; MIps: propensity scores
pooled after multiple imputation; MIpar: propensity score parameters 
pooled after multiple imputation.
	\end{flushleft}
\end{table}%

\begin{table}[htbp]
  \centering
  \caption{Scenario 3: RR=2, $\rho=0.3$ and outcome predictor of missingness and included in the imputation model.}
	\footnotesize
    \begin{tabular}{rcccccccccccccccc}
    \toprule
          & \textbf{Crude} & \textbf{Full } & \textbf{CC*} & \textbf{MP} & \textbf{MIte} & \textbf{MIps} & \textbf{MIpar} \\
    \toprule
    \textbf{RR} &       &       &       &       &       &       &  \\
    Bias  & 0.438 & 0.000 & 0.111 & 0.145 & 0.002 & 0.028 & 0.016 \\
    Variance & 0.004 & 0.006 & 0.013 & 0.007 & 0.006 & 0.006 & 0.006 \\
    Empirical variance & 0.004 & 0.006 & 0.014 & 0.006 & 0.006 & 0.006 & 0.006 \\
    Coverage rate & 0.000 & 0.949 & 0.815 & 0.584 & 0.958 & 0.931 & 0.940 \\
    \textbf{OR} &       &       &       &       &       &       &  \\
    Bias  & 0.747 & 0.002 & 0.130 & 0.227 & 0.004 & 0.047 & 0.026 \\
    Variance & 0.011 & 0.014 & 0.036 & 0.018 & 0.016 & 0.014 & 0.015 \\
    Empirical variance & 0.011 & 0.015 & 0.038 & 0.016 & 0.015 & 0.015 & 0.015 \\
    Coverage rate & 0.000 & 0.947 & 0.887 & 0.622 & 0.958 & 0.930 & 0.942 \\
    \textbf{Risk difference} &       &       &       &       &       &       &  \\
    Bias  & 0.163 & 0.000 & 0.018 & 0.048 & 0.001 & 0.010 & 0.005 \\
    Variance & 0.000 & 0.001 & 0.002 & 0.001 & 0.001 & 0.001 & 0.001 \\
    Empirical variance & 0.000 & 0.001 & 0.002 & 0.001 & 0.001 & 0.001 & 0.001 \\
    Coverage rate  & 0.000 & 0.948 & 0.928 & 0.661 & 0.957 & 0.928 & 0.939 \\
    \bottomrule

    \end{tabular}%
	\begin{flushleft}
	\footnotesize 
	* For complete case analysis, the average sample size is 1076.\\
	RR: relative risk; OR: odds ratio; CC: complete case; MP: missingness pattern;
	MIte:treatment effects pooled after multiple imputation; MIps: propensity scores
pooled after multiple imputation; MIpar: propensity score parameters 
pooled after multiple imputation.
	\end{flushleft}
\end{table}%

\begin{table}[htbp]
  \centering
  \caption{Scenario 4: RR=2, $\rho=0.3$ and outcome independent of missingness but included in the imputation model.}
	\footnotesize
    \begin{tabular}{rcccccccccccccccc}
    \toprule
          & \textbf{Crude} & \textbf{Full } & \textbf{CC*} & \textbf{MP} & \textbf{MIte} & \textbf{MIps} & \textbf{MIpar} \\
    \toprule
    \textbf{RR} &       &       &       &       &       &       &  \\
    Bias  & 0.439 & 0.002 & 0.102 & 0.074 & 0.003 & 0.027 & 0.016 \\
    Variance & 0.004 & 0.006 & 0.015 & 0.007 & 0.006 & 0.006 & 0.006 \\
    Empirical variance & 0.004 & 0.006 & 0.016 & 0.006 & 0.006 & 0.006 & 0.006 \\
    Coverage rate  & 0.000 & 0.950 & 0.850 & 0.868 & 0.958 & 0.931 & 0.941 \\
    \textbf{OR} &       &       &       &       &       &       &  \\
    Bias  & 0.748 & 0.005 & 0.051 & 0.112 & 0.007 & 0.041 & 0.023 \\
    Variance & 0.011 & 0.014 & 0.036 & 0.018 & 0.016 & 0.014 & 0.015 \\
    Empirical variance & 0.011 & 0.015 & 0.038 & 0.016 & 0.015 & 0.015 & 0.015 \\
    Coverage rate  & 0.000 & 0.950 & 0.935 & 0.883 & 0.959 & 0.933 & 0.941 \\
    \textbf{Risk difference} &       &       &       &       &       &       &  \\
    Bias  & 0.163 & 0.001 & -0.014 & 0.023 & 0.001 & 0.008 & 0.004 \\
    Variance & 0.000 & 0.001 & 0.002 & 0.001 & 0.001 & 0.001 & 0.001 \\
    Empirical variance & 0.000 & 0.001 & 0.002 & 0.001 & 0.001 & 0.001 & 0.001 \\
    Coverage rate  & 0.000 & 0.951 & 0.919 & 0.898 & 0.959 & 0.936 & 0.943 \\
    \bottomrule

    \end{tabular}%
	\begin{flushleft}
	\footnotesize 
	* For complete case analysis, the average sample size is 1103.\\
	RR: relative risk; OR: odds ratio; CC: complete case; MP: missingness pattern;
	MIte:treatment effects pooled after multiple imputation; MIps: propensity scores
pooled after multiple imputation; MIpar: propensity score parameters 
pooled after multiple imputation.
	\end{flushleft}
\end{table}%

\begin{table}[htbp]
  \centering
  \caption{Scenario 5: RR=1, $\rho=0.6$ and outcome predictor of missingness and included in the imputation model.}
	\footnotesize
    \begin{tabular}{rcccccccccccccccc}
    \toprule
          & \textbf{Crude} & \textbf{Full } & \textbf{CC*} & \textbf{MP} & \textbf{MIte} & \textbf{MIps} & \textbf{MIpar} \\
    \toprule
    \textbf{RR} &       &       &       &       &       &       &  \\
    Bias  & 0.650 & 0.001 & 0.099 & 0.162 & 0.006 & 0.036 & 0.022 \\
    Variance & 0.006 & 0.009 & 0.027 & 0.012 & 0.009 & 0.009 & 0.009 \\
    Empirical variance & 0.006 & 0.009 & 0.029 & 0.010 & 0.009 & 0.009 & 0.009 \\
    Coverage rate  & 0.000 & 0.945 & 0.886 & 0.686 & 0.955 & 0.928 & 0.937 \\
    \textbf{OR} &       &       &       &       &       &       &  \\
    Bias  & 0.883 & 0.002 & 0.128 & 0.214 & 0.008 & 0.048 & 0.029 \\
    Variance & 0.011 & 0.015 & 0.044 & 0.020 & 0.017 & 0.015 & 0.015 \\
    Empirical variance & 0.012 & 0.016 & 0.047 & 0.017 & 0.016 & 0.016 & 0.016 \\
    Coverage rate  & 0.000 & 0.945 & 0.896 & 0.693 & 0.955 & 0.928 & 0.938 \\
    \textbf{Risk difference} &       &       &       &       &       &       &  \\
    Bias  & 0.169 & 0.001 & 0.023 & 0.040 & 0.002 & 0.009 & 0.006 \\
    Variance & 0.000 & 0.001 & 0.001 & 0.001 & 0.001 & 0.001 & 0.001 \\
    Empirical variance & 0.000 & 0.001 & 0.001 & 0.001 & 0.001 & 0.001 & 0.001 \\
    Coverage rate  & 0.000 & 0.945 & 0.914 & 0.705 & 0.956 & 0.932 & 0.941 \\
    \bottomrule

    \end{tabular}%
	\begin{flushleft}
	\footnotesize 
	* For complete case analysis, the average sample size is 1058.\\
	RR: relative risk; OR: odds ratio; CC: complete case; MP: missingness pattern;
	MIte:treatment effects pooled after multiple imputation; MIps: propensity scores
pooled after multiple imputation; MIpar: propensity score parameters 
pooled after multiple imputation.
	\end{flushleft}
\end{table}%

\begin{table}[htbp]
  \centering
  \caption{Scenario 6: RR=1, $\rho=0.6$ and outcome independent of missingness but included in the imputation model.}
	\footnotesize
    \begin{tabular}{rcccccccccccccccc}
    \toprule
          & \textbf{Crude} & \textbf{Full } & \textbf{CC*} & \textbf{MP} & \textbf{MIte} & \textbf{MIps} & \textbf{MIpar} \\
    \toprule
    \textbf{RR} &       &       &       &       &       &       &  \\
    Bias  & 0.649 & -0.001 & -0.001 & 0.063 & 0.003 & 0.030 & 0.016 \\
    Variance & 0.006 & 0.009 & 0.033 & 0.011 & 0.009 & 0.009 & 0.009 \\
    Empirical variance & 0.006 & 0.009 & 0.035 & 0.010 & 0.009 & 0.009 & 0.009 \\
    Coverage rate  & 0.000 & 0.944 & 0.938 & 0.927 & 0.954 & 0.931 & 0.940 \\
    \textbf{OR} &       &       &       &       &       &       &  \\
    Bias  & 0.881 & 0.000 & 0.002 & 0.084 & 0.005 & 0.040 & 0.022 \\
    Variance & 0.011 & 0.015 & 0.048 & 0.020 & 0.017 & 0.015 & 0.015 \\
    Empirical variance & 0.011 & 0.016 & 0.051 & 0.017 & 0.016 & 0.015 & 0.016 \\
    Coverage rate  & 0.000 & 0.944 & 0.941 & 0.929 & 0.954 & 0.933 & 0.941 \\
    \textbf{Risk difference} &       &       &       &       &       &       &  \\
    Bias  & 0.169 & 0.000 & 0.002 & 0.016 & 0.001 & 0.007 & 0.004 \\
    Variance & 0.000 & 0.001 & 0.001 & 0.001 & 0.001 & 0.001 & 0.001 \\
    Empirical variance & 0.000 & 0.001 & 0.001 & 0.001 & 0.001 & 0.001 & 0.001 \\
    Coverage rate  & 0.000 & 0.945 & 0.938 & 0.934 & 0.953 & 0.936 & 0.941 \\
    \bottomrule

    \end{tabular}%
	\begin{flushleft}
	\footnotesize 
	* For complete case analysis, the average sample size is 1099.\\
	RR: relative risk; OR: odds ratio; CC: complete case; MP: missingness pattern;
	MIte:treatment effects pooled after multiple imputation; MIps: propensity scores
pooled after multiple imputation; MIpar: propensity score parameters 
pooled after multiple imputation.
	\end{flushleft}
\end{table}%

\begin{table}[htbp]
  \centering
  \caption{Scenario 7: RR=2, $\rho=0.6$ and outcome predictor of missingness and included in the imputation model.}
	\footnotesize
    \begin{tabular}{rcccccccccccccccc}
    \toprule
          & \textbf{Crude} & \textbf{Full } & \textbf{CC*} & \textbf{MP} & \textbf{MIte} & \textbf{MIps} & \textbf{MIpar} \\
    \toprule
    \textbf{RR} &       &       &       &       &       &       &  \\
    Bias  & 0.529 & 0.002 & 0.141 & 0.130 & 0.005 & 0.028 & 0.017 \\
    Variance & 0.004 & 0.006 & 0.014 & 0.008 & 0.007 & 0.006 & 0.006 \\
    Empirical variance & 0.004 & 0.006 & 0.014 & 0.007 & 0.006 & 0.006 & 0.006 \\
    Coverage rate  & 0.000 & 0.947 & 0.769 & 0.691 & 0.957 & 0.932 & 0.942 \\
    \textbf{OR} &       &       &       &       &       &       &  \\
    Bias  & 0.924 & 0.005 & 0.148 & 0.210 & 0.009 & 0.047 & 0.028 \\
    Variance & 0.011 & 0.016 & 0.040 & 0.022 & 0.018 & 0.016 & 0.016 \\
    Empirical variance & 0.011 & 0.017 & 0.040 & 0.018 & 0.017 & 0.016 & 0.017 \\
    Coverage rate  & 0.000 & 0.948 & 0.887 & 0.712 & 0.958 & 0.932 & 0.943 \\
    \textbf{Risk difference} &       &       &       &       &       &       &  \\
    Bias  & 0.199 & 0.001 & 0.017 & 0.045 & 0.002 & 0.011 & 0.006 \\
    Variance & 0.000 & 0.001 & 0.002 & 0.001 & 0.001 & 0.001 & 0.001 \\
    Empirical variance & 0.000 & 0.001 & 0.002 & 0.001 & 0.001 & 0.001 & 0.001 \\
    Coverage rate  & 0.000 & 0.945 & 0.937 & 0.731 & 0.957 & 0.932 & 0.942 \\
    \bottomrule

    \end{tabular}%
	\begin{flushleft}
	\footnotesize 
	* For complete case analysis, the average sample size is 1074.\\
	RR: relative risk; OR: odds ratio; CC: complete case; MP: missingness pattern;
	MIte:treatment effects pooled after multiple imputation; MIps: propensity scores
pooled after multiple imputation; MIpar: propensity score parameters 
pooled after multiple imputation.
	\end{flushleft}
\end{table}%

\begin{table}[htbp]
  \centering
  \caption{Scenario 8: RR=2, $\rho=0.6$ and outcome independent of missingness but included in the imputation model.}
	\footnotesize
    \begin{tabular}{rcccccccccccccccc}
    \toprule
          & \textbf{Crude} & \textbf{Full } & \textbf{CC*} & \textbf{MP} & \textbf{MIte} & \textbf{MIps} & \textbf{MIpar} \\
    \toprule
    \textbf{RR} &       &       &       &       &       &       &  \\
    Bias  & 0.531 & 0.002 & 0.138 & 0.057 & 0.005 & 0.027 & 0.016 \\
    Variance & 0.004 & 0.006 & 0.016 & 0.008 & 0.007 & 0.006 & 0.006 \\
    Empirical variance & 0.004 & 0.006 & 0.017 & 0.007 & 0.006 & 0.006 & 0.006 \\
    Coverage rate  & 0.000 & 0.947 & 0.793 & 0.921 & 0.960 & 0.937 & 0.946 \\
    \textbf{OR} &       &       &       &       &       &       &  \\
    Bias  & 0.927 & 0.006 & 0.077 & 0.091 & 0.009 & 0.041 & 0.024 \\
    Variance & 0.011 & 0.016 & 0.039 & 0.021 & 0.018 & 0.016 & 0.016 \\
    Empirical variance & 0.011 & 0.016 & 0.041 & 0.017 & 0.016 & 0.016 & 0.016 \\
    Coverage rate  & 0.000 & 0.949 & 0.935 & 0.924 & 0.959 & 0.938 & 0.944 \\
    \textbf{Risk difference} &       &       &       &       &       &       &  \\
    Bias  & 0.200 & 0.002 & -0.015 & 0.020 & 0.002 & 0.009 & 0.005 \\
    Variance & 0.000 & 0.001 & 0.002 & 0.001 & 0.001 & 0.001 & 0.001 \\
    Empirical variance & 0.000 & 0.001 & 0.002 & 0.001 & 0.001 & 0.001 & 0.001 \\
    Coverage rate  & 0.000 & 0.947 & 0.918 & 0.926 & 0.955 & 0.934 & 0.941 \\
    \bottomrule

    \end{tabular}%
	\begin{flushleft}
	\footnotesize 
	* For complete case analysis, the average sample size is 1099.\\
	RR: relative risk; OR: odds ratio; CC: complete case; MP: missingness pattern;
	MIte:treatment effects pooled after multiple imputation; MIps: propensity scores
pooled after multiple imputation; MIpar: propensity score parameters 
pooled after multiple imputation.
	\end{flushleft}
\end{table}%

\begin{table}[htbp]
  \centering
  \caption{Scenario 9: RR=1, $\rho=0.3$ and outcome predictor of missingness but not included in the imputation model.}
	\footnotesize
    \begin{tabular}{rcccccccccccccccc}
    \toprule
          & \textbf{Crude} & \textbf{Full } & \textbf{CC*} & \textbf{MP} & \textbf{MIte} & \textbf{MIps} & \textbf{MIpar} \\
    \toprule
    \textbf{RR} &       &       &       &       &       &       &  \\
    Bias  & 0.540 & 0.000 & 0.093 & 0.182 & 0.084 & 0.100 & 0.090 \\
    Variance & 0.006 & 0.009 & 0.025 & 0.011 & 0.010 & 0.009 & 0.009 \\
    Empirical variance & 0.006 & 0.009 & 0.026 & 0.010 & 0.008 & 0.008 & 0.008 \\
    Coverage rate  & 0.000 & 0.946 & 0.892 & 0.573 & 0.883 & 0.807 & 0.835 \\
    \textbf{OR} &       &       &       &       &       &       &  \\
    Bias  & 0.724 & 0.001 & 0.121 & 0.239 & 0.111 & 0.132 & 0.119 \\
    Variance & 0.012 & 0.015 & 0.041 & 0.019 & 0.017 & 0.015 & 0.015 \\
    Empirical variance & 0.012 & 0.015 & 0.043 & 0.017 & 0.015 & 0.015 & 0.015 \\
    Coverage rate  & 0.000 & 0.947 & 0.901 & 0.580 & 0.886 & 0.810 & 0.838 \\
    \textbf{Risk difference} &       &       &       &       &       &       &  \\
    Bias  & 0.136 & 0.000 & 0.022 & 0.043 & 0.020 & 0.024 & 0.022 \\
    Variance & 0.000 & 0.000 & 0.001 & 0.001 & 0.001 & 0.001 & 0.001 \\
    Empirical variance & 0.000 & 0.000 & 0.001 & 0.001 & 0.000 & 0.000 & 0.000 \\
    Coverage rate  & 0.000 & 0.947 & 0.920 & 0.598 & 0.892 & 0.820 & 0.849 \\
    \bottomrule

    \end{tabular}%
	\begin{flushleft}
	\footnotesize 
	* For complete case analysis, the average sample size is 1060.\\
	RR: relative risk; OR: odds ratio; CC: complete case; MP: missingness pattern;
	MIte:treatment effects pooled after multiple imputation; MIps: propensity scores
pooled after multiple imputation; MIpar: propensity score parameters 
pooled after multiple imputation.
	\end{flushleft}
\end{table}%

\begin{table}[htbp]
  \centering
  \caption{Scenario 10: RR=1, $\rho=0.3$ and outcome independent of missingness and not included in the imputation model.}
	\footnotesize
    \begin{tabular}{rcccccccccccccccc}
    \toprule
          & \textbf{Crude} & \textbf{Full } & \textbf{CC*} & \textbf{MP} & \textbf{MIte} & \textbf{MIps} & \textbf{MIpar} \\
    \toprule
    \textbf{RR} &       &       &       &       &       &       &  \\
    Bias  & 0.540 & 0.000 & -0.006 & 0.082 & 0.088 & 0.098 & 0.090 \\
    Variance & 0.006 & 0.009 & 0.030 & 0.011 & 0.010 & 0.008 & 0.009 \\
    Empirical variance & 0.006 & 0.009 & 0.032 & 0.009 & 0.008 & 0.008 & 0.008 \\
    Coverage rate  & 0.000 & 0.947 & 0.942 & 0.891 & 0.882 & 0.824 & 0.845 \\
    \textbf{OR} &       &       &       &       &       &       &  \\
    Bias  & 0.724 & 0.000 & -0.005 & 0.108 & 0.116 & 0.129 & 0.119 \\
    Variance & 0.012 & 0.015 & 0.045 & 0.018 & 0.017 & 0.015 & 0.015 \\
    Empirical variance & 0.012 & 0.015 & 0.047 & 0.016 & 0.014 & 0.014 & 0.014 \\
    Coverage rate  & 0.000 & 0.947 & 0.941 & 0.894 & 0.884 & 0.827 & 0.847 \\
    \textbf{Risk difference} &       &       &       &       &       &       &  \\
    Bias  & 0.136 & 0.000 & 0.001 & 0.020 & 0.021 & 0.024 & 0.022 \\
    Variance & 0.000 & 0.000 & 0.001 & 0.001 & 0.001 & 0.000 & 0.000 \\
    Empirical variance & 0.000 & 0.000 & 0.001 & 0.001 & 0.000 & 0.000 & 0.000 \\
    Coverage rate  & 0.000 & 0.948 & 0.940 & 0.903 & 0.893 & 0.835 & 0.855 \\
    \bottomrule

    \end{tabular}%
	\begin{flushleft}
	\footnotesize 
	* For complete case analysis, the average sample size is 1103.\\
	RR: relative risk; OR: odds ratio; CC: complete case; MP: missingness pattern;
	MIte:treatment effects pooled after multiple imputation; MIps: propensity scores
pooled after multiple imputation; MIpar: propensity score parameters 
pooled after multiple imputation.
	\end{flushleft}
\end{table}%

\begin{table}[htbp]
  \centering
  \caption{Scenario 11: RR=2, $\rho=0.3$ and outcome predictor of missingness but not included in the imputation model.}
	\footnotesize
    \begin{tabular}{rcccccccccccccccc}

    \toprule
          & \textbf{Crude} & \textbf{Full } & \textbf{CC*} & \textbf{MP} & \textbf{MIte} & \textbf{MIps} & \textbf{MIpar} \\
    \toprule
    \textbf{RR} &       &       &       &       &       &       &  \\
    Bias  & 0.436 & 0.000 & 0.111 & 0.144 & 0.065 & 0.077 & 0.070 \\
    Variance & 0.004 & 0.006 & 0.013 & 0.007 & 0.006 & 0.006 & 0.006 \\
    Empirical variance & 0.004 & 0.006 & 0.014 & 0.006 & 0.006 & 0.006 & 0.006 \\
    Coverage rate  & 0.000 & 0.946 & 0.818 & 0.583 & 0.893 & 0.825 & 0.854 \\
    \textbf{OR} &       &       &       &       &       &       &  \\
    Bias  & 0.743 & 0.001 & 0.130 & 0.227 & 0.108 & 0.128 & 0.115 \\
    Variance & 0.011 & 0.014 & 0.036 & 0.018 & 0.017 & 0.015 & 0.015 \\
    Empirical variance & 0.011 & 0.015 & 0.038 & 0.017 & 0.015 & 0.015 & 0.015 \\
    Coverage rate  & 0.000 & 0.945 & 0.894 & 0.614 & 0.887 & 0.819 & 0.847 \\
    \textbf{Risk difference} &       &       &       &       &       &       &  \\
    Bias  & 0.162 & 0.000 & 0.018 & 0.048 & 0.024 & 0.028 & 0.025 \\
    Variance & 0.000 & 0.001 & 0.002 & 0.001 & 0.001 & 0.001 & 0.001 \\
    Empirical variance & 0.000 & 0.001 & 0.002 & 0.001 & 0.001 & 0.001 & 0.001 \\
    Coverage rate  & 0.000 & 0.944 & 0.923 & 0.646 & 0.886 & 0.822 & 0.851 \\
    \bottomrule

    \end{tabular}%
	\begin{flushleft}
	\footnotesize 
	* For complete case analysis, the average sample size is 1075.\\
	RR: relative risk; OR: odds ratio; CC: complete case; MP: missingness pattern;
	MIte:treatment effects pooled after multiple imputation; MIps: propensity scores
pooled after multiple imputation; MIpar: propensity score parameters 
pooled after multiple imputation.
	\end{flushleft}
\end{table}%

\begin{table}[htbp]
  \centering
  \caption{Scenario 12: RR=2, $\rho=0.3$ and outcome independent of missingness and not included in the imputation model.}
	\footnotesize
    \begin{tabular}{rcccccccccccccccc}
    \toprule
          & \textbf{Crude} & \textbf{Full } & \textbf{CC*} & \textbf{MP} & \textbf{MIte} & \textbf{MIps} & \textbf{MIpar} \\
    \toprule
    \textbf{RR} &       &       &       &       &       &       &  \\
    Bias  & 0.437 & 0.002 & 0.099 & 0.073 & 0.069 & 0.079 & 0.073 \\
    Variance & 0.004 & 0.006 & 0.015 & 0.007 & 0.006 & 0.006 & 0.006 \\
    Empirical variance & 0.004 & 0.006 & 0.016 & 0.006 & 0.006 & 0.006 & 0.006 \\
    Coverage rate  & 0.000 & 0.947 & 0.860 & 0.875 & 0.879 & 0.824 & 0.843 \\
    \textbf{OR} &       &       &       &       &       &       &  \\
    Bias  & 0.745 & 0.004 & 0.047 & 0.110 & 0.115 & 0.126 & 0.117 \\
    Variance & 0.011 & 0.014 & 0.036 & 0.018 & 0.017 & 0.015 & 0.015 \\
    Empirical variance & 0.011 & 0.015 & 0.038 & 0.016 & 0.014 & 0.014 & 0.015 \\
    Coverage rate  & 0.000 & 0.947 & 0.936 & 0.887 & 0.877 & 0.829 & 0.846 \\
    \textbf{Risk difference} &       &       &       &       &       &       &  \\
    Bias  & 0.162 & 0.001 & -0.015 & 0.023 & 0.026 & 0.027 & 0.025 \\
    Variance & 0.000 & 0.001 & 0.002 & 0.001 & 0.001 & 0.001 & 0.001 \\
    Empirical variance & 0.000 & 0.001 & 0.002 & 0.001 & 0.001 & 0.001 & 0.001 \\
    Coverage rate  & 0.000 & 0.946 & 0.913 & 0.901 & 0.879 & 0.836 & 0.855 \\
    \bottomrule

    \end{tabular}%
	\begin{flushleft}
	\footnotesize 
	* For complete case analysis, the average sample size is 1103.\\
	RR: relative risk; OR: odds ratio; CC: complete case; MP: missingness pattern;
	MIte:treatment effects pooled after multiple imputation; MIps: propensity scores
pooled after multiple imputation; MIpar: propensity score parameters 
pooled after multiple imputation.
	\end{flushleft}
\end{table}%

\begin{table}[htbp]
  \centering
  \caption{Scenario 13: RR=1, $\rho=0.6$ and outcome predictor of missingness but not included in the imputation model.}
	\footnotesize
    \begin{tabular}{rcccccccccccccccc}

    \toprule
          & \textbf{Crude} & \textbf{Full } & \textbf{CC*} & \textbf{MP} & \textbf{MIte} & \textbf{MIps} & \textbf{MIpar} \\
    \toprule
    \textbf{RR} &       &       &       &       &       &       &  \\
    Bias  & 0.650 & 0.001 & 0.096 & 0.161 & 0.061 & 0.082 & 0.070 \\
    Variance & 0.006 & 0.009 & 0.027 & 0.012 & 0.010 & 0.009 & 0.009 \\
    Empirical variance & 0.006 & 0.009 & 0.029 & 0.010 & 0.009 & 0.009 & 0.009 \\
    Coverage rate  & 0.000 & 0.939 & 0.896 & 0.683 & 0.919 & 0.858 & 0.882 \\
    \textbf{OR} &       &       &       &       &       &       &  \\
    Bias  & 0.883 & 0.002 & 0.124 & 0.214 & 0.081 & 0.109 & 0.094 \\
    Variance & 0.011 & 0.015 & 0.044 & 0.020 & 0.017 & 0.015 & 0.015 \\
    Empirical variance & 0.012 & 0.016 & 0.046 & 0.017 & 0.015 & 0.015 & 0.015 \\
    Coverage rate  & 0.000 & 0.938 & 0.905 & 0.690 & 0.921 & 0.861 & 0.885 \\
    \textbf{Risk difference} &       &       &       &       &       &       &  \\
    Bias  & 0.169 & 0.001 & 0.022 & 0.040 & 0.015 & 0.020 & 0.017 \\
    Variance & 0.000 & 0.001 & 0.001 & 0.001 & 0.001 & 0.001 & 0.001 \\
    Empirical variance & 0.000 & 0.001 & 0.001 & 0.001 & 0.001 & 0.001 & 0.001 \\
    Coverage rate  & 0.000 & 0.939 & 0.925 & 0.703 & 0.925 & 0.867 & 0.890 \\
    \bottomrule

    \end{tabular}%
	\begin{flushleft}
	\footnotesize 
	* For complete case analysis, the average sample size is 1058.\\
	RR: relative risk; OR: odds ratio; CC: complete case; MP: missingness pattern;
	MIte:treatment effects pooled after multiple imputation; MIps: propensity scores
pooled after multiple imputation; MIpar: propensity score parameters 
pooled after multiple imputation.
	\end{flushleft}
\end{table}%

\begin{table}[htbp]
  \centering
  \caption{Scenario 14: RR=1, $\rho=0.6$ and outcome independent of missingness and not included in the imputation model.}
	\footnotesize
    \begin{tabular}{rcccccccccccccccc}

    \toprule
          & \textbf{Crude} & \textbf{Full } & \textbf{CC*} & \textbf{MP} & \textbf{MIte} & \textbf{MIps} & \textbf{MIpar} \\
    \toprule
    \textbf{RR} &       &       &       &       &       &       &  \\
    Bias  & 0.650 & -0.001 & -0.001 & 0.063 & 0.063 & 0.080 & 0.070 \\
    Variance & 0.006 & 0.009 & 0.033 & 0.011 & 0.010 & 0.009 & 0.009 \\
    Empirical variance & 0.006 & 0.009 & 0.034 & 0.009 & 0.008 & 0.008 & 0.008 \\
    Coverage rate  & 0.000 & 0.952 & 0.937 & 0.931 & 0.928 & 0.870 & 0.893 \\
    \textbf{OR} &       &       &       &       &       &       &  \\
    Bias  & 0.882 & 0.000 & 0.002 & 0.083 & 0.084 & 0.106 & 0.093 \\
    Variance & 0.011 & 0.015 & 0.048 & 0.020 & 0.017 & 0.015 & 0.015 \\
    Empirical variance & 0.012 & 0.015 & 0.051 & 0.016 & 0.014 & 0.014 & 0.014 \\
    Coverage rate  & 0.000 & 0.953 & 0.939 & 0.935 & 0.929 & 0.873 & 0.896 \\
    \textbf{Risk difference} &       &       &       &       &       &       &  \\
    Bias  & 0.169 & 0.000 & 0.002 & 0.015 & 0.016 & 0.020 & 0.017 \\
    Variance & 0.000 & 0.001 & 0.001 & 0.001 & 0.001 & 0.001 & 0.001 \\
    Empirical variance & 0.000 & 0.001 & 0.001 & 0.001 & 0.001 & 0.000 & 0.000 \\
    Coverage rate  & 0.000 & 0.953 & 0.938 & 0.938 & 0.932 & 0.882 & 0.904 \\
    \bottomrule

    \end{tabular}%
	\begin{flushleft}
	\footnotesize 
	* For complete case analysis, the average sample size is 1099.\\
	RR: relative risk; OR: odds ratio; CC: complete case; MP: missingness pattern;
	MIte:treatment effects pooled after multiple imputation; MIps: propensity scores
pooled after multiple imputation; MIpar: propensity score parameters 
pooled after multiple imputation.
	\end{flushleft}
\end{table}%

\begin{table}[htbp]
  \centering
  \caption{Scenario 15: RR=2, $\rho=0.6$ and outcome predictor of missingness but not included in the imputation model.}
	\footnotesize
    \begin{tabular}{rcccccccccccccccc}
    \toprule
          & \textbf{Crude} & \textbf{Full } & \textbf{CC*} & \textbf{MP} & \textbf{MIte} & \textbf{MIps} & \textbf{MIpar} \\
    \toprule
    \textbf{RR} &       &       &       &       &       &       &  \\
    Bias  & 0.529 & 0.000 & 0.139 & 0.129 & 0.048 & 0.064 & 0.056 \\
    Variance & 0.004 & 0.006 & 0.014 & 0.008 & 0.007 & 0.006 & 0.006 \\
    Empirical variance & 0.004 & 0.006 & 0.014 & 0.006 & 0.006 & 0.006 & 0.006 \\
    Coverage rate  & 0.000 & 0.953 & 0.763 & 0.696 & 0.929 & 0.878 & 0.901 \\
    \textbf{OR} &       &       &       &       &       &       &  \\
    Bias  & 0.924 & 0.001 & 0.144 & 0.207 & 0.081 & 0.107 & 0.092 \\
    Variance & 0.011 & 0.016 & 0.039 & 0.021 & 0.018 & 0.017 & 0.017 \\
    Empirical variance & 0.011 & 0.016 & 0.041 & 0.018 & 0.016 & 0.016 & 0.016 \\
    Coverage rate  & 0.000 & 0.953 & 0.897 & 0.716 & 0.927 & 0.872 & 0.894 \\
    \textbf{Risk difference} &       &       &       &       &       &       &  \\
    Bias  & 0.199 & 0.001 & 0.016 & 0.045 & 0.018 & 0.024 & 0.021 \\
    Variance & 0.000 & 0.001 & 0.002 & 0.001 & 0.001 & 0.001 & 0.001 \\
    Empirical variance & 0.000 & 0.001 & 0.002 & 0.001 & 0.001 & 0.001 & 0.001 \\
    Coverage rate  & 0.000 & 0.948 & 0.936 & 0.737 & 0.925 & 0.869 & 0.894 \\
    \bottomrule

    \end{tabular}%
	\begin{flushleft}
	\footnotesize 
	* For complete case analysis, the average sample size is 1074.\\
	RR: relative risk; OR: odds ratio; CC: complete case; MP: missingness pattern;
	MIte:treatment effects pooled after multiple imputation; MIps: propensity scores
pooled after multiple imputation; MIpar: propensity score parameters 
pooled after multiple imputation.
	\end{flushleft}
\end{table}%

\begin{table}[htbp]
  \centering
  \caption{Scenario 16: RR=2, $\rho=0.6$ and outcome independent of missingness and not included in the imputation model.}
	\footnotesize
    \begin{tabular}{rcccccccccccccccc}

    \toprule
          & \textbf{Crude} & \textbf{Full } & \textbf{CC*} & \textbf{MP} & \textbf{MIte} & \textbf{MIps} & \textbf{MIpar} \\
    \toprule
    \textbf{RR} &       &       &       &       &       &       &  \\
    Bias  & 0.530 & 0.001 & 0.139 & 0.057 & 0.051 & 0.065 & 0.058 \\
    Variance & 0.004 & 0.006 & 0.016 & 0.008 & 0.007 & 0.006 & 0.006 \\
    Empirical variance & 0.004 & 0.006 & 0.018 & 0.007 & 0.006 & 0.006 & 0.006 \\
    Coverage rate  & 0.000 & 0.944 & 0.795 & 0.918 & 0.923 & 0.869 & 0.891 \\
    \textbf{OR} &       &       &       &       &       &       &  \\
    Bias  & 0.926 & 0.003 & 0.077 & 0.090 & 0.086 & 0.105 & 0.093 \\
    Variance & 0.011 & 0.016 & 0.039 & 0.021 & 0.018 & 0.016 & 0.017 \\
    Empirical variance & 0.011 & 0.017 & 0.042 & 0.018 & 0.016 & 0.017 & 0.017 \\
    Coverage rate  & 0.000 & 0.945 & 0.928 & 0.925 & 0.921 & 0.875 & 0.894 \\
    \textbf{Risk difference} &       &       &       &       &       &       &  \\
    Bias  & 0.199 & 0.001 & -0.015 & 0.019 & 0.019 & 0.023 & 0.020 \\
    Variance & 0.000 & 0.001 & 0.002 & 0.001 & 0.001 & 0.001 & 0.001 \\
    Empirical variance & 0.000 & 0.001 & 0.002 & 0.001 & 0.001 & 0.001 & 0.001 \\
    Coverage rate  & 0.000 & 0.945 & 0.912 & 0.927 & 0.919 & 0.877 & 0.894 \\
    \bottomrule

    \end{tabular}%
	\begin{flushleft}
	\footnotesize
	* For complete case analysis, the average sample size is 1099.\\
	RR: relative risk; OR: odds ratio; CC: complete case; MP: missingness pattern;
	MIte:treatment effects pooled after multiple imputation; MIps: propensity scores
pooled after multiple imputation; MIpar: propensity score parameters 
pooled after multiple imputation.
	\end{flushleft}
\end{table}%

\clearpage
\subsection*{6.2: Impact of the sample size}
\begin{table}[htbp]
  \centering
  \caption{Results for one scenario (RR=2, $\rho=0.6$ and outcome predictor of missingness and included in the imputation model)
	with $n=500$.}
	\footnotesize
    \begin{tabular}{rcccccccccccccccc}
    
            \toprule
          & \textbf{Crude} & \textbf{Full } & \textbf{CC*} & \textbf{MP} & \textbf{MIte} & \textbf{MIps} & \textbf{MIpar} \\
    \toprule
    \textbf{RR} &       &       &       &       &       &       &  \\
    Bias  & 0.442 & 0.007 & 0.110 & 0.153 & 0.010 & 0.038 & 0.024 \\
    Variance & 0.017 & 0.022 & 0.050 & 0.029 & 0.026 & 0.022 & 0.023 \\
    Empirical variance & 0.018 & 0.024 & 0.059 & 0.027 & 0.025 & 0.024 & 0.025 \\
    Coverage rate  & 0.065 & 0.940 & 0.887 & 0.855 & 0.955 & 0.939 & 0.943 \\
    \textbf{OR} &       &       &       &       &       &       &  \\
    Bias  & 0.753 & 0.015 & 0.145 & 0.244 & 0.020 & 0.066 & 0.041 \\
    Variance & 0.044 & 0.057 & 0.139 & 0.077 & 0.065 & 0.058 & 0.058 \\
    Empirical variance & 0.045 & 0.061 & 0.168 & 0.071 & 0.063 & 0.062 & 0.063 \\
    Coverage rate  & 0.047 & 0.942 & 0.913 & 0.865 & 0.952 & 0.934 & 0.937 \\
    \textbf{Risk difference} &       &       &       &       &       &       &  \\
    Bias  & 0.162 & 0.002 & 0.021 & 0.051 & 0.003 & 0.013 & 0.008 \\
    Variance & 0.002 & 0.003 & 0.007 & 0.004 & 0.003 & 0.003 & 0.003 \\
    Empirical variance & 0.002 & 0.003 & 0.009 & 0.004 & 0.003 & 0.003 & 0.003 \\
    Coverage rate  & 0.046 & 0.934 & 0.902 & 0.869 & 0.947 & 0.928 & 0.929 \\
    \bottomrule

    \end{tabular}%
			\begin{flushleft}
	\footnotesize 
	* For complete case analysis, the average sample size is 269.\\
	RR: relative risk; OR: odds ratio; CC: complete case; MP: missingness pattern;
	MIte:treatment effects pooled after multiple imputation; MIps: propensity scores
pooled after multiple imputation; MIpar: propensity score parameters 
pooled after multiple imputation.
	\end{flushleft}
\end{table}%

\clearpage

\subsection*{6.3: Impact of the missingness rate}
\begin{table}[htbp]
  \centering
  \caption{Results for one scenario (RR=2, $\rho=0.6$ and outcome predictor of missingness and included in the imputation model)
	with 10\% of data missing for $X_1$ and $X_3$.}
	\footnotesize
    \begin{tabular}{rcccccccccccccccc}
    
    \toprule
          & \textbf{Crude} & \textbf{Full } & \textbf{CC*} & \textbf{MP} & \textbf{MIte} & \textbf{MIps} & \textbf{MIpar} \\
    \toprule
    \textbf{RR} &       &       &       &       &       &       &  \\
    Bias  & 0.529 & 0.000 & 0.066 & 0.061 & 0.001 & 0.009 & 0.005 \\
    Variance & 0.004 & 0.006 & 0.007 & 0.008 & 0.006 & 0.006 & 0.006 \\
    Empirical variance & 0.004 & 0.006 & 0.008 & 0.007 & 0.006 & 0.006 & 0.006 \\
    Coverage rate  & 0.000 & 0.946 & 0.874 & 0.907 & 0.951 & 0.945 & 0.947 \\
    \textbf{OR} &       &       &       &       &       &       &  \\
    Bias  & 0.924 & 0.002 & 0.073 & 0.098 & 0.003 & 0.015 & 0.009 \\
    Variance & 0.011 & 0.016 & 0.020 & 0.021 & 0.017 & 0.016 & 0.016 \\
    Empirical variance & 0.012 & 0.017 & 0.021 & 0.018 & 0.017 & 0.017 & 0.017 \\
    Coverage rate  & 0.000 & 0.944 & 0.913 & 0.904 & 0.949 & 0.945 & 0.946 \\
    \textbf{Risk difference} &       &       &       &       &       &       &  \\
    Bias  & 0.199 & 0.001 & 0.010 & 0.021 & 0.001 & 0.004 & 0.002 \\
    Variance & 0.000 & 0.001 & 0.001 & 0.001 & 0.001 & 0.001 & 0.001 \\
    Empirical variance & 0.000 & 0.001 & 0.001 & 0.001 & 0.001 & 0.001 & 0.001 \\
    Coverage rate  & 0.000 & 0.942 & 0.933 & 0.905 & 0.948 & 0.943 & 0.944 \\
    \bottomrule

    \end{tabular}%
	\begin{flushleft}
* For complete case analysis, the average sample size is 1609.\\	
RR: relative risk; OR: odds ratio; CC: complete case; MP: missingness pattern;
	MIte:treatment effects pooled after multiple imputation; MIps: propensity scores
pooled after multiple imputation; MIpar: propensity score parameters 
pooled after multiple imputation.
	\end{flushleft}
\end{table}%

\begin{table}[htbp]
  \centering
  \caption{Results for one scenario (RR=2, $\rho=0.6$ and outcome predictor of missingness and included in the imputation model)
	with 60\% of data missing for $X_1$ and $X_3$.}
	\footnotesize
    \begin{tabular}{rcccccccccccccccc}
    
        \toprule
          & \textbf{Crude} & \textbf{Full } & \textbf{CC*} & \textbf{MP} & \textbf{MIte} & \textbf{MIps} & \textbf{MIpar} \\
    \toprule
    \textbf{RR} &       &       &       &       &       &       &  \\
    Bias  & 0.530 & 0.001 & 0.184 & 0.175 & 0.010 & 0.063 & 0.037 \\
    Variance & 0.004 & 0.006 & 0.074 & 0.008 & 0.008 & 0.006 & 0.006 \\
    Empirical variance & 0.004 & 0.006 & 0.101 & 0.007 & 0.007 & 0.006 & 0.007 \\
    Coverage rate  & 0.000 & 0.945 & 0.802 & 0.490 & 0.962 & 0.869 & 0.916 \\
    \textbf{OR} &       &       &       &       &       &       &  \\
    Bias  & 0.925 & 0.004 & 0.184 & 0.285 & 0.018 & 0.105 & 0.060 \\
    Variance & 0.011 & 0.016 & 0.208 & 0.025 & 0.021 & 0.016 & 0.017 \\
    Empirical variance & 0.012 & 0.017 & 0.285 & 0.022 & 0.019 & 0.017 & 0.018 \\
    Coverage rate  & 0.000 & 0.944 & 0.889 & 0.544 & 0.963 & 0.863 & 0.916 \\
    \textbf{Risk difference} &       &       &       &       &       &       &  \\
    Bias  & 0.199 & 0.001 & 0.010 & 0.062 & 0.004 & 0.023 & 0.013 \\
    Variance & 0.000 & 0.001 & 0.011 & 0.001 & 0.001 & 0.001 & 0.001 \\
    Empirical variance & 0.000 & 0.001 & 0.014 & 0.001 & 0.001 & 0.001 & 0.001 \\
    Coverage rate  & 0.000 & 0.945 & 0.867 & 0.578 & 0.963 & 0.862 & 0.913 \\
    \bottomrule

    \end{tabular}%
	\begin{flushleft}
	\footnotesize 
	* For complete case analysis, the average sample size is 399.\\
	RR: relative risk; OR: odds ratio; CC: complete case; MP: missingness pattern;
	MIte:treatment effects pooled after multiple imputation; MIps: propensity scores
pooled after multiple imputation; MIpar: propensity score parameters 
pooled after multiple imputation.
	\end{flushleft}
\end{table}%

\clearpage
\subsection*{6.5: Number of imputed datasets}
\begin{table}[htbp]
  \centering
  \caption{Bias of the log(RR), its variance and coverage rate for the 3 MI approaches according the number $M$ of imputed datasets for one scenario
	(RR=2, $\rho=0.6$, outcome predictor of missingness and included in the imputation model).}
    \begin{tabular}{rccccccccc}
    \toprule
          & \multicolumn{3}{c}{\textbf{MIte}} & \multicolumn{3}{c}{\textbf{MIps}} & \multicolumn{3}{c}{\textbf{MIpar}} \\
    \toprule
          & \textit{M=5} & \textit{M=10} & \textit{M=20} & \textit{M=5} & \textit{M=10} & \textit{M=20} & \textit{M=5} & \textit{M=10} & \textit{M=20} \\
    Bias  & 0.002 & 0.005 & 0.003 & 0.026 & 0.028 & 0.031 & 0.015 & 0.017 & 0.018 \\
    Variance & 0.006 & 0.007 & 0.006 & 0.006 & 0.006 & 0.006 & 0.006 & 0.006 & 0.006 \\
    Empirical variance & 0.006 & 0.006 & 0.006 & 0.006 & 0.006 & 0.006 & 0.006 & 0.006 & 0.006 \\
    Coverage rate  & 0.957 & 0.957 & 0.953 & 0.931 & 0.932 & 0.924 & 0.940 & 0.942 & 0.937 \\
    \bottomrule
		\end{tabular}
		~\\
		\vspace{0.2cm}
	\footnotesize  MIte:
treatment effects pooled after multiple imputation; MIps: propensity scores
pooled after multiple imputation; MIpar: propensity score parameters 
pooled after multiple imputation. 
  \label{tab_M}%
\end{table}%

\clearpage
\subsection*{6.5: Partially observed outcome and treatment indicator}
\begin{table}[htbp]
  \centering
  \caption{Results for one scenario (RR=2, $\rho=0.6$ and outcome predictor of missingness and included in the imputation model)
	with 30\% of data missing for $X_1$, $X_3$, the outcome $Y$ and the treatment indicator $Z$.}
	\footnotesize
    \begin{tabular}{rcccccccccccccccc}
    
    \toprule
          & \textbf{Crude} & \textbf{Full } & \textbf{CC*} & \textbf{MIte} \\
    \toprule
    \textbf{RR} &       &       &       &  \\
    Bias  & 0.530 & 0.002 & 0.207 & 0.000 \\
    Variance & 0.004 & 0.006 & 0.032 & 0.014 \\
    Empirical variance & 0.004 & 0.006 & 0.034 & 0.012 \\
    Coverage rate  & 0.000 & 0.949 & 0.770 & 0.956 \\
    \textbf{OR} &       &       &       &  \\
    Bias  & 0.926 & 0.005 & 0.164 & -0.002 \\
    Variance & 0.011 & 0.016 & 0.078 & 0.034 \\
    Empirical variance & 0.011 & 0.016 & 0.083 & 0.031 \\
    Coverage rate  & 0.000 & 0.947 & 0.906 & 0.952 \\
    \textbf{Risk difference} &       &       &       &  \\
    Bias  & 0.200 & 0.001 & -0.002 & -0.001 \\
    Variance & 0.000 & 0.001 & 0.004 & 0.002 \\
    Empirical variance & 0.000 & 0.001 & 0.004 & 0.002 \\
    Coverage rate  & 0.000 & 0.950 & 0.923 & 0.953 \\
    \bottomrule

    \end{tabular}%
	\begin{flushleft}
* For complete case analysis, the average sample size is 635.\\
RR: relative risk; OR: odds ratio; CC: complete case; MP: missingness pattern;
	MIte:treatment effects pooled after multiple imputation; MIps: propensity scores
pooled after multiple imputation; MIpar: propensity score parameters 
pooled after multiple imputation.
	\end{flushleft}
\end{table}%

\end{document}